\documentclass[aps,tightenlines,twocolumn,prd,nofootinbib,
superscriptaddress,showpacs,preprintnumbers,eqsecnum,floatfix]{revtex4-2}
\def\bea{\begin{eqnarray}}
\def\be{\begin{equation}}
\def\ee{\end{equation}}
\def\eea{\end{eqnarray}}
\def\bal{\begin{align}}
\def\eal{\end{align}}

\usepackage[usenames,dvipsnames]{color}
\usepackage{graphics}
\usepackage{graphicx}
\usepackage{amsmath}
\usepackage{amssymb}
\usepackage{bbold}
\usepackage{bm}
\usepackage{slashed}
\usepackage{float}
\usepackage{psfrag}
\usepackage{footnote} 
\usepackage{xcolor}
\usepackage{textcomp}
\usepackage{cancel}
\usepackage{dcolumn}
\usepackage{caption}
\usepackage{subcaption}
\begin{document}

\title{Heavy and heavy-light mesons with arbitrary spin and parity in the Covariant Spectator Theory}

\author{Alfred Stadler}
\email{stadler@uevora.pt}
\affiliation{Departamento de F\'isica, Universidade de \'Evora, 7000-671 \'Evora, Portugal}
\affiliation{Laborat\'orio de Instrumenta\c{c}\~ao e F\'isica Experimental de Part\'iculas---LIP, Avenida Professor Gama Pinto, 2, 1649-003 Lisboa, Portugal}

\author{Elmar P. Biernat}
\email{elmar.biernat@tecnico.ulisboa.pt}
\affiliation{C$^2$\!TN and Departamento de Engenharia e Ci\^encias Nucleares, Instituto Superior T\'ecnico, Universidade de Lisboa, Campus Tecnol\'ogico e Nuclear, 2695-066 Bobadela, Portugal}
\affiliation{Laborat\'orio de Instrumenta\c{c}\~ao e F\'isica Experimental de Part\'iculas---LIP, Avenida Professor Gama Pinto, 2, 1649-003 Lisboa, Portugal}

\date{\today}
\begin{abstract}
This work generalizes the one-channel Covariant Spectator Theory (CST) formalism to describe quark-antiquark mesons of arbitrary spin-parity $J^P$. We also improve the quark-antiquark interaction kernel by incorporating the momentum dependence of the strong coupling. Within this framework, we perform global fits to the masses of experimentally established heavy and heavy-light mesons with $J^P=0^\pm, 1^\pm, 2^\pm,$ and $3^\pm$. With only eight adjustable parameters, the model yields an excellent global description of the observed quark-antiquark spectrum and predicts both unmeasured states and likely $J^P$ assignments for states with  unconfirmed quantum numbers. In particular, our results support the identifications of the recently observed $B_c(1P)^+$ candidates as $0^+$, $1^+$, and $2^+$ states, and the $B_c^{*+}$ as the lowest $1^-$ state in the bottom-charm sector, as well as the $D_{s1}(2933)^+$ as axial-vector state in the charm-strange sector.
\end{abstract}

\pacs{11.10.St, 14.40.Pq, 12.39.Pn, 03.65.Ge}
\keywords{}

\maketitle


\section{Introduction}\label{sec1}

The Particle Data Group (PDG) meson summary table~\cite{ParticleDataGroup:2026aaa} currently lists around 70 established quark-antiquark states that contain at least one heavy quark. Several of these mesons, and also new states in the bottom-charm and charm-strange sectors, have been measured only recently by LHCb, ATLAS, BESIII, and Belle II~\cite{ATLAS:2026ubk, LHCb:2026jk,LHCb:2025uce, LHCb:2025ubr, LHCb:2024vfz, LHCb:2020gnv, ZHANG202490, Brambilla_2011}. While the majority of the listed states are spin-0 and spin-1 mesons, 14 are classified as higher-spin (tensor) states with total angular momentum $J\geq2$. 

Various theoretical frameworks have been developed to describe the properties of mesons with higher spin. These include constituent-quark models~\cite{Godfrey1985, PhysRevD.72.054026, LAKHINA2007159, Llanes_Estrada_2000, LLANESESTRADA2002303, PhysRevD.64.114004, PhysRevD.84.034006, PhysRevD.79.094004,PhysRevD.91.094004, Liu2016}, covariant Dyson-Schwinger/Bethe-Salpeter (DS-BS)~\cite{LLEWELLYNSMITH1969521, 10.1143/PTP.46.897, PhysRevD.51.2347, PhysRevD.75.036007,Koll_2000, PhysRevC.49.494, Krassnigg2011,Fischer_2014,Popovici_2014,Hilger_2017,Hagel:2026cl} and quasi-potential approaches~\cite{Zeng:1995xr,Ebert2009,  PhysRevC.49.494}, models based on two-body Dirac equations~\cite{PhysRevD.82.094020,PhysRevD.87.034021}, the basis light-front quantization~\cite{PhysRevD.96.016022, Leitao:2017esb},
 chiral models~\cite{Jafarzade2021, Jafarzade2022}, unitarized models \cite{van_Beveren_2021}, effective field theories~\cite{RevModPhys.77.1423, PhysRevD.81.094001}, and lattice QCD~\cite{Gayer_2025, PhysRevD.81.034508, PhysRevD.85.054509,PhysRevD.72.094507,PhysRevD.85.114509,PhysRevD.90.034510,PhysRevD.77.034501,Liu2012,Dudek_2013,PhysRevD.83.111502}. 
 
 The QCD-motivated quark model by Godfrey and Isgur (GI)~\cite{Godfrey1985} has played a particularly influential role. It describes mesons as quark-antiquark pairs interacting via a one-gluon-exchange plus linear-confinement potential, with relativistic effects incorporated in a semiquantitative manner. Meson states with total angular momentum $J$, parity $P$, and (for quarkonia) charge-conjugation parity $C$ are constructed by coupling the total quark-antiquark spin $S$ and the orbital angular momentum $L$, with $P=(-1)^{1+L}$ and $C=(-1)^{L+S}$. The GI model was the first to provide a unified description of the masses and couplings of light, strange, and heavy mesons of total angular momentum up to $J = 5$. 

The description of mesons with arbitrary $J^{P(C)}$ in fully covariant approaches, such as the DS-BS approach and its three-dimensional reductions differs substantially from the nonrelativistic treatment. These approaches are rooted in quantum field theory and imply the presence of relativistic partial-wave components with opposite orbital parity, in addition to the dominant nonrelativistic components familiar from constituent-quark models. While relativistic components are expected to be small for heavy quarkonia and bottom-charmed mesons, they can become important in heavy-light systems. The DS-BS approach is very
successful in light quark systems, but has difficulties for heavy-light systems and for higher excited states. 

The Covariant Spectator Theory (CST) is based on the Gross equation (GE)~\cite{Gross:1969rv,PhysRevC.26.2203,mass_function_paper,Biernat:2014xaa}, a three-dimensional reduction of the Bethe-Salpeter equation. It has previously been applied to spin-0 and spin-1 mesons~\cite{Gross:1991te,Gross:1991pk,Gross:1994he,Savkli:1999me}, it is particularly suited for heavy-light systems, and most recently, the one-channel GE (1CGE) was used to calculate the masses and vertex functions of all heavy and heavy-light  mesons with $J\leq 1$~\cite{Leit_o_2017,PhysRevD.96.074007, Leitao:2017bds, Stadler:2018hjv}. In these studies, a covariant quark-antiquark interaction with Lorentz scalar, pseudoscalar, and vector structures was employed which reduces to the Cornell potential in the non-relativistic limit, similar to the GI model. The adjustable parameters of the model consist of three interaction strengths, the quark masses, and the relative weight between Lorentz scalar$+$pseudoscalar and vector confinement. Global least-square fits to the experimental spectrum, ranging from charmed mesons up to bottomonium with $J\leq1$, yielded a remarkably good description of the data. Notably, even when only the subset of pseudoscalar states were included in the fit, the predicted vector, scalar, and axial-vector meson masses achieved nearly the same level of accuracy as in fits in which additional states were included. This demonstrates that the Lorentz-covariant structure of the kernel correctly encodes the spin-dependence of the interaction. 

The main objective of the present work is to extend the $q\bar{q}$ 1CGE for $J>1$ and to calculate the corresponding higher-spin states in the heavy and heavy-light meson spectrum.  An accurate description of tensor mesons as highly excited states is particularly important in studies of confinement, since they are expected to be more sensitive than lower lying states to details of the confining interaction compared to one-gluon exchange. Therefore, these states provide a valuable testing ground for different mechanisms of confinement and their Lorentz-covariant structure. It also enables the predictions of not yet observed states and provides guidance for assigning $J^P$ quantum numbers to experimentally observed states that are not yet firmly established or have uncertain quark content.    

The needed generalization of the $q\bar{q}$ 1CGE implies mainly the derivation of the general form of the CST $q\bar{q}$ wave function for arbitrary spin and parity. Up to now, only the special cases of pseudoscalar, scalar, vector and axial-vector meson wave functions have been derived in CST. A new formulation of the structure of the wave function for arbitrary spin and parity is an important step on the way to a CST description of all $q\bar{q}$ mesons. 

In addition to generalizing the spin-structure of the meson wave functions, we also we refine the interaction kernel. In previous CST calculations, the strong coupling $\alpha_\text{s}$ was treated as a constant. Here, we make our OGE interaction kernel more realistic by incorporating the momentum dependence of $\alpha_\text{s}$ and investigate its consequences.       

This paper is organized as follows: In Sec.~\ref{sec2}, we present the 1CSE Gross equation and the interaction kernel. Section~\ref{sec3} introduces the spherical-tensor basis used to solve the resulting system of coupled CST equations. The numerical results are discussed in Sec.~\ref{sec5}. Finally, Sec.~\ref{sec6} summarizes our findings and presents our conclusions. Two appendices collect longer formulas and  derivations.
 
\section{The one-channel CST formalism}\label{sec2}
\subsection{The Gross equation for mesons}\label{sec2.1}

The one-channel CST equation~\cite{Gross:1969rv}, traditionally called the Gross equation (1CGE), 
\begin{eqnarray}
\Gamma(\hat p_1,p_2)=-\int \frac{\mathrm d^3 k_1}{(2\pi)^3}\frac{m_1}{E_{1k}} \mathcal V(\hat p_1,\hat k_1) \Lambda(\hat k_1)\Gamma(\hat k_1,k_2)  S(k_2)\, , 
\nonumber\\
\label{eq:1CSE}
\end{eqnarray}
describes a meson as a bound state of a quark and an antiquark.
For a derivation of this equation starting from the Bethe-Salpeter equation, see, for instance, Ref.~\cite{PhysRevD.96.074007}. 
In (\ref{eq:1CSE}), $\Gamma(\hat p_1,p_2)$ is the CST meson vertex function which determines the spin-parity quantum numbers $J^P$ of the meson; $\hat p_1=(E_{1p},\boldsymbol p)$ and $\hat k_1=(E_{1k},\boldsymbol k)$ are the external and internal four-momenta  of quark 1 (a ``\^{ }''  indicates that a four-momentum is on mass shell) with constituent mass $m_1$ and energies $E_{1p}=\sqrt{m_1^2+\boldsymbol p^2}$ and $E_{1k}=\sqrt{m_1^2+\boldsymbol k^2}$, respectively; $p_2=\hat p_1-P$ and $k_2=\hat k_1-P$ are the external and internal four-momenta of quark 2, respectively; $P=(\mu,\boldsymbol 0)$ is the four-momentum of the meson with mass $\mu$ in its rest frame;
$\Lambda(\hat k_1)=\left(m_1+\hat {\slashed {k}}_1\right)/2m_1$ is the positive-energy projector of quark~1;
\begin{eqnarray}
S(k_2)=\frac {m_2+\slashed {k}_2}{m_2^2-k_2^2-\mathrm i \epsilon}
\end{eqnarray}
is the propagator of quark 2 with constant\footnote{In this work, we assume, for simplicity, constant instead of dynamical (momentum-dependent) quark masses. } constituent mass $m_2$; and $\mathcal V(\hat p_1,\hat k_1)$ is the CST interaction kernel~\cite{PhysRevD.96.074007}
\begin{widetext}

\begin{eqnarray}
\mathcal V(\hat p_1,\hat k_1)&=&\sum_{K=\text {L,G,C}}\Theta_1^K\otimes \Theta_2^K V_K(\hat p_1,\hat k_1) \nonumber\\&=&\boldsymbol 1_1\otimes\boldsymbol 1_2 \, V_{\text L}(\hat p_1,\hat k_1)-\gamma^\mu_1\otimes\gamma_{\mu2} \left[ V_{\text G}(\hat p_1,\hat k_1)+ V_{\text C}(\hat p_1,\hat k_1)\right]\,.\label{eq:kernel}
\end{eqnarray}
Here $V_{\text L}$ is a covariant generalization of the linear
confining potential
\begin{eqnarray}
V_{\text L}(\hat p_1,\hat k_1)&=&-8\sigma \pi \left[\left(\frac{1}{(\hat p_1-\hat k_1)^4}-\frac{1}{\Lambda_{\rm L}^4+(\hat p_1-\hat k_1)^4}\right)\right.\nonumber\\&&\left.-\frac{E_{1p}}{m_1} (2\pi)^3\delta^3(\boldsymbol p-\boldsymbol k) \int \frac{\mathrm d^3 k_1'}{(2\pi)^3}\frac{m_1}{E_{1k'}}\left(\frac{1}{(\hat p_1-\hat k_1')^4}-\frac{1}{\Lambda_{\rm L}^4+(\hat p_1-\hat k_1')^4}\right)\right]\,,
\end{eqnarray} 
$V_{\text G}$ is a one-gluon exchange in Feynman gauge,
\begin{eqnarray}
V_{\text G}(\hat p_1,\hat k_1)&=&-4\pi \alpha_{\rm s} \left((\hat p_1-\hat k_1)^2\right) \left(\frac{1}{(\hat p_1-\hat k_1)^2}-\frac{1}{(\hat p_1-\hat k_1)^2-\Lambda_{\rm G}^2}\right)\,,
\end{eqnarray} 
\end{widetext}
and $V_{\text C}$ is a covariant generalization of an (in coordinate space) constant potential,
\begin{eqnarray}
V_{\text C}(\hat p_1,\hat k_1)&=&\frac{E_{1k}}{m_1} (2\pi)^3 C\delta^3(\boldsymbol p-\boldsymbol k) \,.
\end{eqnarray} 
The coupling strengths $\sigma$ and $C$ are adjustable parameters and $\alpha_{\rm s} \left((\hat p_1-\hat k_1)^2\right)$ is the strong running coupling 
\begin{eqnarray}
	\alpha_{\rm s}(q^2)=\frac{1}{\beta_0 \mathrm {ln}\left(\frac{-q^2}{\Lambda^2_{\rm QCD}}+\tau\right)}\, ,
\end{eqnarray}
where $N_f$ is the number of active flavors and $\beta_0=\frac{33-2N_f}{12\pi}$. The IR regulator $\tau$ is fixed by specifying the value of $\alpha_{\rm s}(q^2)$ at $q^2=0$. The remaining parameter, $\Lambda_{\rm QCD}$, is determined by requiring that $\alpha_{\rm s}(q^2)$ reproduces the experimental value at $q^2=-M_Z^2$. 

We regularize the UV-behavior of linear-confining and OGE kernels via Pauli-Villars subtraction with cut-off parameters $\Lambda_{\rm L}=\lambda_{\rm L}m_1$ and  $\Lambda_{\rm G}=\lambda_{\rm G}m_1$, respectively. The dimensionless cut-off parameters $\lambda_{\rm L}$ and $\lambda_{\rm G}$ together with the coupling strengths $\sigma$, $\alpha_{\rm s}(0)$, and $C$ are adjustable fit parameters.      
\subsection{Positive- and negative-energy channels}\label{sec2.2}

For solving Eq.~(\ref{eq:1CSE}), it is convenient to decompose the equation into its positive- and negative-energy channels. This is achieved by taking matrix elements of the vertex functions and interaction vertices between four-spinors $u^\rho$ with $\rho=+,-$. These $\rho$-spinors are related to the standard $u$ and $v$ Dirac spinors in the Bjorken-Drell convention. For quark 1 we define the positive- and negative-energy spinors
\begin{eqnarray}
 u^+_1(\boldsymbol p,\lambda_1)\equiv u_1(\boldsymbol p,\lambda_1)=N_{1p} 
 \left(\begin{array}{c}
   1\\
   \boldsymbol\sigma\cdot \hat {\boldsymbol p}\, \tilde p_1
 \end{array}\right)\otimes \chi^{}_{\lambda_1}
 \nonumber\\\label{eq:uspinor}
\end{eqnarray}
and 
\begin{eqnarray}
u^-_1(\boldsymbol p,\lambda)\equiv v_1(-\boldsymbol p,\lambda_1)=N_{1p}
 \left(\begin{array}{c}
   -\boldsymbol\sigma\cdot \hat {\boldsymbol p} \, \tilde p_1\\1
   
 \end{array}\right) \otimes \chi^{}_{\lambda_1} \,.
 \nonumber\\\label{eq:vspinor}
\end{eqnarray}
For quark 2 the corresponding definitions are
\begin{eqnarray}
	u^+_2(\boldsymbol p,\lambda_2)&\equiv& C\bar v_2^\intercal(\boldsymbol p,\lambda_2)\nonumber\\&=& N_{2p} 
	\left(\begin{array}{c}
		1\\
		\boldsymbol\sigma\cdot \hat {\boldsymbol p} \, \tilde p_2
	\end{array}\right)\otimes (\mathrm i\,\sigma^2 \chi^{}_{\lambda_2})\label{eq:up2spinor}\nonumber\\
\end{eqnarray}
and 

\begin{eqnarray}
	u^-_2(\boldsymbol p,\lambda_2)&\equiv&  C \bar u_2^\intercal(-\boldsymbol p,\lambda_2)\nonumber\\&=&N_{2p}
	\left(\begin{array}{c}
		-\boldsymbol\sigma\cdot \hat {\boldsymbol p}\, \tilde p_2\\1
		
	\end{array}\right) \otimes (-\mathrm i\,\sigma^2 \chi^{}_{\lambda_2}) \,.\nonumber\\\label{eq:um2spinor}
\end{eqnarray}
Here we use the notation $N_{ip}=\sqrt{(E_{ip}+m_i)/2m_i}$, $\tilde p_i=p/(E_{ip}+m_i)$,
and the Pauli spinors
\begin{eqnarray}
\chi^{}_{\lambda_i}= \left(\begin{array}{cc}
     \frac12 + \lambda_i  \\
     \frac12 - \lambda_i 
\end{array}\right) \, \quad \text{with}\quad \lambda_i=\pm\frac12\,. \label{eq:chispinor}
\end{eqnarray}
which describe spin-$1/2$ states quantized along the $z$-direction. The charge-conjugation matrix is
$C=-\mathrm i\,\gamma^0\gamma^2$. 
With the choice $\mathrm i\, \sigma^2 \chi^{}_{\lambda_2}$ for the Pauli spinor of quark 2, matrix elements of spherical tensor operators reduce to sums of Clebsch-Gordan coefficients multiplied by spherical harmonics in the Condon-Shortley convention.   

The projector and propagator under the integral in~(\ref{eq:1CSE}) can be expressed in terms of the $\rho$-spinors:
\begin{eqnarray}
\Lambda(\hat k_1)= \sum_{\lambda_1'=\pm \frac12} u_1^+(\boldsymbol k,\lambda_1') \bar{u}_1^+(\boldsymbol k, \lambda_1' )
\end{eqnarray}
and
\begin{eqnarray}
S(k_2)=\frac{m_2}{E_{2k}}\sum_{\rho'=\pm} \sum_{\lambda_2'=\pm \frac12} \frac{u_2^{\rho'}(\boldsymbol k,\lambda_2') \bar{u}_2^{\rho'}(\boldsymbol k, \lambda_2' )}{\rho'E_{2k}- E_{1k}+\mu-\mathrm i\,\epsilon}\,.\nonumber\\\label{eq:Sexpand}
\end{eqnarray}
Substituting these in (\ref{eq:1CSE}) and multiplying from the left with $\bar u_1^+(\boldsymbol p,\lambda_1)$ and from the right with $u_2^\rho(\boldsymbol p,\lambda_2)$ yields two coupled equations for the projected amplitudes:
\begin{widetext}
\begin{eqnarray}
\bar u_1^+(\boldsymbol p,\lambda_1)\Gamma(\hat p_1,p_2)u_2^\rho(\boldsymbol p,\lambda_2)=&-&\sum_{K\rho'\lambda_1'\lambda_2'}\int \frac{\mathrm d^3 k_1}{(2\pi)^3}\frac{m_1}{E_{1k}} \frac{m_2}{E_{2k}}V_K(\hat p_1,\hat k_1) \bar u_1^+(\boldsymbol p,\lambda_1) \Theta_1^K  u_1^+(\boldsymbol k,\lambda_1')  \nonumber\\&&\times   \frac{\bar{u}_1^+(\boldsymbol k, \lambda_1' ) \Gamma(\hat k_1,k_2)u_2^{\rho'}(\boldsymbol k,\lambda_2') }{\rho'E_{2k}- E_{1k}+\mu-\mathrm i\,\epsilon} \bar{u}_2^{\rho'}(\boldsymbol k, \lambda_2' )  \Theta_2^K u_2^\rho(\boldsymbol p,\lambda_2) \, ,
\label{eq:1CSEa} 
\end{eqnarray}
with $\rho=\pm$.
The CST wave functions are then defined as 
\begin{eqnarray}
\Psi^{+\rho}_{\lambda_1\lambda_2}(\boldsymbol p) &=&\sqrt{\frac{m_1 m_2}{E_{1p}E_{2p}}} \frac{ \bar{u}_1^+(\boldsymbol p, \lambda_1 ) \Gamma(\hat p_1,p_2)u_2^\rho(\boldsymbol p, \lambda_2 ) }{\rho E_{2p}-  E_{1p}+\mu-\mathrm i\,\epsilon} \, ,
\label{eq:CSTwfs}
\end{eqnarray} 
and the Dirac spinor matrix elements of the interaction vertices are
\begin{eqnarray}
\Theta^{K\rho\rho'}_{\lambda_i\lambda_i'}(\boldsymbol p,\boldsymbol k)&=&
\bar u_i^\rho(\boldsymbol p,\lambda_i) \Theta_i^K  u_i^{\rho'}(\boldsymbol k,\lambda_i')\, .
\label{eq:Theta}
\end{eqnarray} 
With these definitions, (\ref{eq:1CSEa}) can be written as 
\begin{eqnarray}
\left(\rho E_{2p}-  E_{1p}+\mu\right)\Psi^{+\rho}_{\lambda_1\lambda_2}(\boldsymbol p)=&-&\sum_{K\rho'\lambda_1'\lambda_2'}\int \frac{\mathrm d^3 k_1}{(2\pi)^3}N_{12}(p,k) V_K(\hat p_1,\hat k_1)  \Theta^{K++}_{\lambda_1\lambda_1'}(\boldsymbol p,\boldsymbol k)    \Psi^{+\rho'}_{\lambda_1'\lambda_2'}(\boldsymbol k) \Theta^{K\rho'\rho}_{\lambda_2'\lambda_2}(\boldsymbol k,\boldsymbol p)\label{eq:1CSEaa}\, ,
\end{eqnarray}
where \begin{eqnarray}
N_{12}(p,k)=\frac{m_1m_2}{\sqrt{E_{1p}E_{2p}E_{1k}E_{2k}}}\,.
\end{eqnarray}
\end{widetext}
\section{Spherical-tensor basis}\label{sec3}
\subsection{$2\times 2$ spin matrices and partial waves}
To solve Eqs.~(\ref{eq:1CSEaa}), it is convenient to expand the wave functions $\Psi^{+\rho}_{\lambda_1\lambda_2}(\boldsymbol p)$ in a basis of eigenfunctions of the total orbital angular momentum $L$ and total spin $S$ of the quark-antiquark system. Although neither $L$ nor $S$ are conserved in a relativistic framework, this expansion is useful for comparing our results to nonrelativistic approaches, which classify states according to $L$ and $S$.
To implement this, we express the wave functions (\ref{eq:CSTwfs}) and interaction vertex
matrix elements (\ref{eq:Theta}) in terms of matrix elements
of the two-component Pauli spinors $\chi^{}_{\lambda_i}$ of (\ref{eq:chispinor}), using the Dirac spinor
representation of (\ref{eq:uspinor}) and (\ref{eq:vspinor}). This leads to 
\begin{eqnarray}
&&\Theta^{K\rho\rho'}_{\lambda_1\lambda_1'}(\boldsymbol p,\boldsymbol k)=N_{1p}N_{1k}\chi^{\dag}_{\lambda_1} M^{K\rho\rho'}_1 (\boldsymbol p,\boldsymbol k)\chi^{}_{\lambda_1'}\,,\label{eq:ThetaK1}
\\&&\Theta^{K\rho\rho'}_{\lambda_2\lambda_2'}(\boldsymbol p,\boldsymbol k)=N_{2p}N_{2k}\chi^{\dag}_{\lambda_2}(-\mathrm i \, \sigma^2) M^{K\rho\rho'}_2 (\boldsymbol p,\boldsymbol k)\mathrm i \, \sigma^2\chi^{}_{\lambda_2'}\,.\nonumber\\\label{eq:ThetaK2}
\end{eqnarray} 
Similarly, the wave function can be expanded as
\begin{eqnarray}
	\Psi^{+\rho}_{\lambda_1\lambda_2}(\boldsymbol p) &=&\sum_{j=1,2}\psi_{j}^\rho (p)\chi_{\lambda_1}^{\dag} K^{\rho}_j (\hat{\boldsymbol p})\mathrm i\, \sigma^2\chi^{}_{\lambda_2}\,.
	\label{eq:Psip}	
\end{eqnarray}
The $2\times 2$ matrices $M^{K}$ depend on the Lorentz structure of the interaction vertex (labeled by the upper index $K$), while the operators $ K^{\rho}_j$ are four linearly independent spherical tensor operators. These were originally introduced in Ref.~\cite{PhysRevD.96.074007} for the special cases $J=0$ and $J=1$, noting that only the two operators  $K^{-\rho}_1$ and $K^\rho_2$ exist for $J=0$. In this work, we extend the definition of the operators $ K^{\rho}_j$  to arbitrary total angular momentum $J$.
Together with the associated radial wave function $\psi_{j}^\rho (p)$, they describe the four possible orbital and spin configurations 
$(L,S)$ for a given total angular momentum $J\geq 1$: one spin-singlet and three spin-triplet cases, namely

\begin{enumerate}
			\item[(i)]  $L=J$, $S=0$ ,
			\item[(ii)] $L=J$, $S=1$ ,
			\item[(iii)] $L=J-1$, $S=1$ ,
			\item[(iv)]  $L=J+1$, $S=1$ .
\end{enumerate} 
 Configurations (i) and (ii) have parity $P=(-1)^{J}$ (natural parity), while configurations (iii) and (iv) have parity $P=(-1)^{J+1}$ (unnatural parity). Note that for $J=0$, only the two configurations (i) and (iv) exist.  
 
To assign the $K^{\rho}_j$ operators to these cases, one must distinguish between natural-parity mesons ($P=(-1)^J$), such as scalar ($0^+$), vector ($1^-$), and tensor ($2^+, 3^-$, \ldots) and unnatural-parity mesons ($P=(-1)^{J+1}$), such as pseudoscalar ($0^-$), axial-vector ($1^+$), and axial-tensor ($2^-, 3^+$, \ldots) mesons. 
Consider first the $\Psi^{++}$ component of the wave function obtained by multiplying $\Gamma(\hat p_1,p_2)$ with $\bar u^+$ and $u^+$. Since $u^+$ has positive intrinsic parity, $\Psi^{++}$ and hence the corresponding operators $K^+_j$ inherit the parity of the vertex function. Conversely, $\Psi^{+-}$, obtained by multiplying $\Gamma(\hat p_1,p_2)$ with $\bar u^+$ and $u^-$ carries opposite parity because $u^-$ has negative intrinsic parity. Accordingly, the operators $K^-_j$ are opposite in parity to the vertex function. 

Thus, for natural-parity mesons, $K^+_1$ and $K^+_2$ correspond to cases (i) and (ii) while $K^-_1$ and $K^-_2$ correspond to cases (iii) and (iv). For unnatural-parity mesons, the assignments of $K^-_j$ and $K^+_j$ are interchanged.
The four operators $K^\rho_j$ can be identified with irreducible spherical tensors of total angular momentum $J$, denoted $T^{(L,S)}_{m_J}$, constructed by coupling orbital angular momentum $L$ and spin $S$. Explicit expressions of the four $T^{(L,S)}_{m_J}$ in terms of the three-momentum direction $\hat {\boldsymbol p}$, the Pauli matrices $\boldsymbol {\sigma}$ and the spin-1 polarization vectors $\boldsymbol {\xi}$ are derived in Appendix~\ref{eq:derivationKrhoj} and given in Eqs.~(\ref{eq:T0J})-(\ref{eq:T1Jp1}). The correspondence between $T^{(L,S)}_{m_J}$ and $K^\rho_j$ operators is
\begin{eqnarray}
	&&K_1^{-\rho}(\hat{\boldsymbol p})\equiv\sqrt 2\sqrt {4\pi} T^{(J,0)}_{m_J}(\hat{\boldsymbol p})(-\mathrm i \,\sigma^2)\,,\\
	&&K_2^{-\rho}(\hat{\boldsymbol p})\equiv\sqrt 2\sqrt {4\pi}T^{(J,1)}_{m_J}(\hat{\boldsymbol p})(-\mathrm i \,\sigma^2)\,,\\
	&&K^{\rho}_1(\hat{\boldsymbol p})\equiv\sqrt 2 \sqrt {4\pi} T^{(J-1,1)}_{m_J}(\hat{\boldsymbol p}) (-\mathrm i \,\sigma^2)\,,\\
	&&K^{\rho}_2(\hat{\boldsymbol p})	\equiv\sqrt 2\sqrt {4\pi}T^{(J+1,1)}_{m_J}(\hat{\boldsymbol p})(-\mathrm i \,\sigma^2)\,,	
\end{eqnarray}
with $\rho=-$ for natural-parity and $\rho=+$ for unnatural-parity meson states. For the special case $J=0$, this reduces to
\begin{eqnarray}&&K_1^{-\rho}(\hat{\boldsymbol p})\equiv \sqrt 2\sqrt {4\pi} T^{(0,0)}_{0}(\hat{\boldsymbol p})(-\mathrm i \,\sigma^2)\,,\\
	&&K_2^{-\rho}(\hat{\boldsymbol p})\equiv0\,,\\
	&&K^{\rho}_1(\hat{\boldsymbol p})\equiv 0\,,\\
	&&K^{\rho}_2(\hat{\boldsymbol p})	\equiv\sqrt 2\sqrt {4\pi}T^{(1,1)}_{0}(\hat{\boldsymbol p})(-\mathrm i \,\sigma^2)\,.
\end{eqnarray}

Since the $K^\rho_j$ operators appear between Pauli spinors in expression (\ref{eq:Psip}) for $\Psi^{+\rho}_{\lambda_1\lambda_2}(\boldsymbol p)$, the relevant quantities are matrix elements of $T^{(L,S)}_{m_J}$ between Pauli spinors of quark 1 and quark 1, which evaluate to
\begin{eqnarray}
	&&\chi^{\dag}_{\lambda_1}	T^{(L,S)}_{m_J}(\hat{\boldsymbol p}) \chi^{}_{\lambda_2}\nonumber\\&&=\sum_{m_L=-L}^L\sum_{m_S=-S}^S Y_{Lm_L} (\hat{\boldsymbol p})C_{\frac12\lambda_1 \frac12 \lambda_2 }^{Sm_S}  C_{Lm_LS m_S}^{Jm_J}\,  . \nonumber\\ \label{eq:TSLSHCGC}
\end{eqnarray}
Here we have used expression~(\ref{eq:spintensor}) together with the orthogonality of the Pauli spinors. Substituting (\ref{eq:TSLSHCGC}) in  (\ref{eq:Psip}) allows to express the CST wave functions $\Psi^{+\rho}_{\lambda_1\lambda_2}(\boldsymbol p)$ in terms of partial waves:
\begin{eqnarray}
	\Psi^{+\rho}_{\lambda_1\lambda_2}(\boldsymbol p)&=& \sqrt 2\sqrt {4\pi}
	\sum_{L=J\pm 1} \sum_{m_L=-L}^L\sum_{m_S=-1}^1\psi_{L,1} (p)  Y_{Lm_L} (\hat{\boldsymbol p})\nonumber\\ &&\times C_{\frac12\lambda_1 \frac12 \lambda_2 }^{1m_S}C_{Lm_L1 m_S}^{Jm_J}\,,\label{eq:PsiprhoCGSH}\\
	\Psi^{+\,-\!\rho}_{\lambda_1\lambda_2}(\boldsymbol p) &=& \sqrt 2\sqrt {4\pi}
	\sum_{S=0, 1} \sum_{m_L=-J}^J\sum_{m_S=-S}^S\psi_{J,S} (p) Y_{Jm_L} (\hat{\boldsymbol p})\nonumber\\ &&\times C_{\frac12\lambda_1 \frac12 \lambda_2 }^{Sm_S} C_{Jm_LS m_S}^{Jm_J}\, ,
	\label{eq:PsipmrhoCGSH}
\end{eqnarray}
with 
\begin{eqnarray}& \psi_{J-1,1} (p)=\psi^\rho_1 (p)\,,\quad &\psi_{J+1,1} (p)= \psi^\rho_2 (p)\,,\\ 
&\psi_{J,0} (p)=\psi^{-\rho}_1 (p) \,,\quad & \psi_{J,1} (p)=\psi^{-\rho}_2 (p)\,,
\end{eqnarray}
where $\rho=-$ for natural-parity and $\rho=+$ for unnatural-parity mesons.

\subsection{1CGE in spherical-tensor basis}
To cast the 1CGE into a form suitable for numerical solution, we insert (\ref{eq:ThetaK1}), (\ref{eq:ThetaK2}) and (\ref{eq:Psip}) for the interaction vertex matrix elements and CST wave functions, respectively, into Eq.~(\ref{eq:1CSEaa}). Applying the completeness relation of the Pauli spinors
\begin{eqnarray}
	\sum_{\lambda_1} \chi^{}_{\lambda_1} \chi_{\lambda_1}^{\dag}=	\sum_{\lambda_2} \mathrm i\, \sigma^2\chi^{}_{\lambda_2} (\mathrm i\, \sigma^2\chi^{}_{\lambda_2})^{\dag} =\mathbf 1\,
\end{eqnarray} then yields
\begin{eqnarray}
	&&	\left(\rho E_{2p}-  E_{1p}+\mu\right)\sum_{j=1,2}\psi_{j}^\rho (p) K^{\rho}_j (\hat{\boldsymbol p})\nonumber\\&=&-\sum_{K}\int \frac{\mathrm d^3 k_1}{(2\pi)^3}N
	(p,k) V_K(\hat p_1,\hat k_1)  M^{K++}_1(\boldsymbol p,\boldsymbol k) \nonumber\\&&\times\sum_{\rho'=\pm, j'=1,2}\psi_{j'}^{\rho'} (k) K^{\rho'}_{j'} (\hat{\boldsymbol k})  M^{K\rho'\rho}_{2}(\boldsymbol k,\boldsymbol p)\, ,
	\label{eq:1CSEaaa}
\end{eqnarray}
where \begin{eqnarray}
	N(p,k)=N_{1p} N_{1k}N_{2k} N_{2p}N_{12}(p,k)\,.
\end{eqnarray} 
These expressions can be simplified by using the fact that the CST interaction kernel in the 1CGE depends only on the magnitudes  $p\equiv |\boldsymbol p|$ and $k\equiv |\boldsymbol k|$ of the three-momenta $\boldsymbol p$ and $\boldsymbol k$, and on the angle $z\equiv \hat{\boldsymbol p}\cdot \hat{\boldsymbol k}$ between them, i.e.,

\begin{eqnarray}
	V_K(\hat p_1,\hat k_1)\equiv V^K(p,k,z)\,.
\end{eqnarray} 
Using this property, it can be shown that the left- and right-hand sides of Eq.~(\ref{eq:1CSEaa}) have the same tensorial structure. Hence, the equation can be written in the form
\begin{eqnarray}
	&&	\left(\rho E_{2p}-  E_{1p}+\mu\right)\sum_{j}\psi_{j}^\rho (p) K^{\rho}_j (\hat{\boldsymbol p}) \nonumber\\&&=-\sum_{K}\int \frac{\mathrm d^3 k_1}{(2\pi)^3}N
	(p,k) V_K(p,k,z) \sum_{j}  K_j^{\rho}(\hat{\boldsymbol p})\nonumber\\&&\quad \times\sum_{j'\rho'} 
	A^{K\rho\rho'}_{jj'}(p,k,z)\psi_{j'}^{\rho'}(k)\,,\label{eq:1CSEaaa}
\end{eqnarray}
where the new functions $A^{K\rho\rho'}_{jj'}(p,k,z)$ introduced here can be determined explicitly for each $J^{P}$ state.
Because tensors $K^\rho_j$ and $K^{\rho'}_{j'}$ with different $L\neq L'$ and/or $S\neq S'$---or, equivalently, different indices $j\neq j'$ and/or $\rho\neq \rho'$---are orthogonal (see Eq.~(\ref{eq:Kortonorm})), one can project out the coefficients of each tensor in (\ref{eq:1CSEaaa}). After multiplying with $K^{\rho\dag}_j$, integrating over the angles, and taking the trace, one arrives at the system of coupled eigenvalue equations
\begin{eqnarray}
	&&	\left(-\rho E_{2p}+  E_{1p}\right)\psi_{j}^\rho (p)\nonumber\\&&-\sum_{K}\int \frac{\mathrm d^3 k_1}{(2\pi)^3}N
	(p,k) V_K(p,k,z) \sum_{j'\rho'} 
	A^{K\rho\rho'}_{jj'}(p,k,z)\psi_{j'}^{\rho'}(k)  \nonumber\\&&=\mu\psi_{j}^\rho (p)\,.\label{eq:1CSEaaaa}
\end{eqnarray}
This linear eigenvalue equation determines the meson masses $\mu$ and the associated eigenvectors, which are the radial partial wave functions $\psi_{j}^\rho (p)$. Solving this equation thus simultaneously yields the ground state and all excited states for a given $J^{P}$. Once the partial wave functions are obtained, the meson vertex function $\Gamma(\hat p_1,p_2)$ can be constructed, as discussed in Appendix~\ref{sec4}. The vertex function is useful in many applications, such as the calculation of meson form factors~\cite{Biernat:2014b,PhysRevD.92.076011}.

\section{Numerical results and discussion}\label{sec5}

\begin{table}[tb]
	\centering
	\caption{Parameter values of the three models with $N_f=2$.}
	\label{tab1}
	\begin{tabular}{l|cc|cccc|cc}
		\hline
		\hline
		& \multicolumn{2}{|c|}{Coupling} & \multicolumn{4}{c}{ }&\multicolumn{2}{|c}{Cut-off} \\
		Fitted   & \multicolumn{2}{|c|}{strengths} & \multicolumn{4}{c}{Quark masses [GeV]}&\multicolumn{2}{|c}{parameters} \\
		states& $\sigma$ [GeV$^2$]& $\alpha_{\rm s}(0)$&$m_b $&$m_c $&$m_s $&$m_q $&$\lambda_{\rm L}$&$\lambda_{\rm G}$ \\
		\hline
		10 & 0.2158 & 0.4186 & 4.794 & 1.441 & 0.274 & 0.133 & 1.219 & 1.786 \\
		33 & 0.1785 & 0.5074 & 4.852 & 1.508 & 0.343 & 0.185 & 2.812 & 2.266 \\
		49 & 0.1755 & 0.5225 & 4.859 & 1.517 & 0.353 & 0.197 & 2.903 & 2.243 \\
		\hline
		\hline
	\end{tabular}
\end{table}

\begin{figure*}[htb]
\centering
\centering
\includegraphics[width=\linewidth]{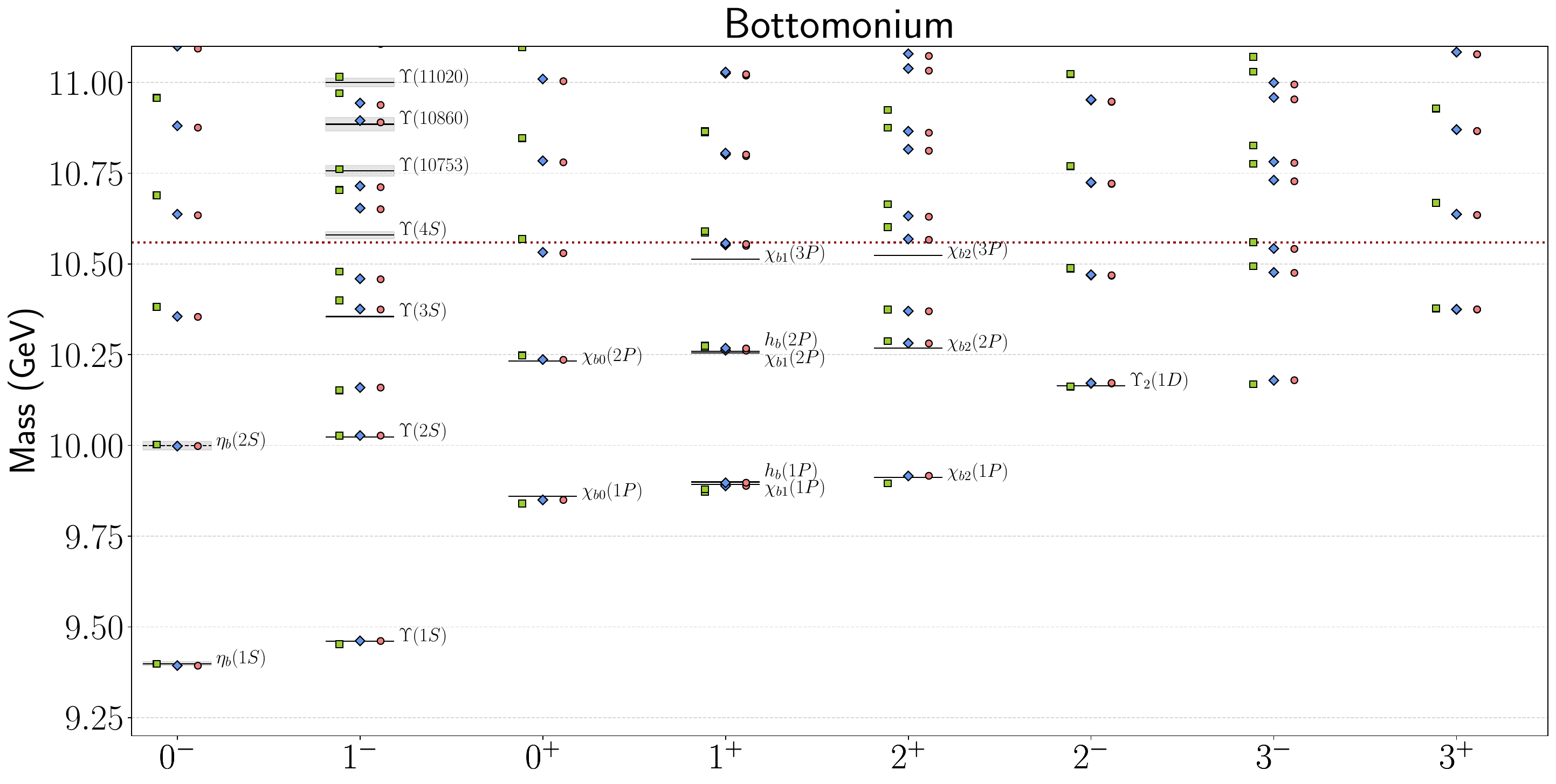}
	\caption{(Color online) Masses of bottomonium with $J^P$ from $0^\pm$ to $3^\pm$. The results of fits to pseudoscalar (green squares), non-axial (blue diamonds), and to states of all $J^P$ (red circles), are compared with established (solid lines) and unconfirmed (dashed lines) experimental masses, with the grey shading displaying the width. The red dotted line indicates the open-bottom threshold.}
\label{fig:bb}
\end{figure*}

\begin{table*}[ht]
\caption{Bottomonium masses. All states are coupled, and the dominant state with quantum numbers $n\, ^{2S+1}\!L_J$ is the one with the highest probability. }
\label{tab:bb}
\begin{minipage}[t]{0.45\textwidth}
\vspace{0pt}
\begin{ruledtabular}
 \begin{tabular}{r|ccc}
  & Dominant & & \\
Meson  & state & This work & PDG  \\
\hline
$J^P=0^-$\\
\hline
$\eta_b(1S)$  & $1\, ^1\!S_0$  &  9.393 & 9.3987 \\
 $\eta_b(2S)$ & $2\, ^1\!S_0$  &  9.998 & 9.999 \\
              & $3\, ^1\!S_0$  & 10.354 &  \\
              & $4\, ^1\!S_0$  & 10.634 &  \\
              & $5\, ^1\!S_0$  & 10.876 &  \\
              & $6\, ^1\!S_0$  & 11.094 &  \\
 \hline
 $J^P=1^-$\\
\hline
$\Upsilon(1S)$ & $1\, ^3\!S_1$ &  9.461 & 9.46040 \\
 $\Upsilon(2S)$ & $2\, ^3\!S_1$ & 10.027 & 10.0234 \\
                & $1\, ^3\!D_1$ & 10.159 & \\
 $\Upsilon(3S)$ & $3\, ^3\!S_1$ & 10.375 & 10.3551 \\
                & $2\, ^3\!D_1$ & 10.458 & \\
 $\Upsilon(4S)$ & $4\, ^3\!S_1$ & 10.651 & 10.5794 \\
 $\Upsilon(10753)$ & $3\, ^3\!D_1$ & 10.712 & 10.7566 \\
 $\Upsilon(10860)$ & $5\, ^3\!S_1$ & 10.890 & 10.8852 \\
                & $4\, ^3\!D_1$ & 10.938 &  \\
 $\Upsilon(11020)$ & $6\, ^3\!S_1$ & 11.106 & 11.000 \\
 \hline
$J^P=0^+$\\
 \hline
 $\chi_{b0}(1P)$ & $1\, ^3\!P_0$ &  9.850 & 9.85944 \\
 $\chi_{b0}(2P)$ & $2\, ^3\!P_0$ & 10.236 & 10.2325 \\
                 & $3\, ^3\!P_0$ & 10.530 & \\
                 & $4\, ^3\!P_0$ & 10.780 & \\
                 & $5\, ^3\!P_0$ & 11.004 & \\
\hline
$J^P=1^+$\\
 \hline
  $\chi_{b1}(1P)$ & $1\, ^3\!P_1$ &  9.889 & 9.89278 \\
 $h_b(1P)$       & $1\, ^1\!P_1$ &  9.897 & 9.8993 \\
 $\chi_{b1}(2P)$ & $2\, ^3\!P_1$ & 10.261 & 10.25546 \\
 $h_b(2P)$       & $2\, ^1\!P_1$ & 10.267 & 10.2598 \\
 $\chi_{b1}(3P)$ & $3\, ^3\!P_1$ & 10.550 & 10.5134 \\
                 & $3\, ^1\!P_1$ & 10.555 & \\
                 & $4\, ^3\!P_1$ & 10.797 & \\
                 & $4\, ^1\!P_1$ & 10.801 & \\
\end{tabular} 
\end{ruledtabular}
\end{minipage}
\hfill
\begin{minipage}[t]{0.45\textwidth}
\vspace{0pt}
\begin{ruledtabular}
 \begin{tabular}{r|ccc}
  & Dominant & & \\
Meson  & state & This work & PDG  \\
\hline
 $J^P=2^+$\\
 \hline
 $\chi_{b2}(1P)$ & $1\, ^3\!P_2$ &  9.916 & 9.91221 \\
 $\chi_{b2}(2P)$ & $2\, ^3\!P_2$ & 10.281 & 10.26865 \\
                 & $1\, ^3\!F_2$ & 10.370 &  \\
 $\chi_{b2}(3P)$ & $3\, ^3\!P_2$ & 10.567 & 10.5240 \\
                 & $2\, ^3\!F_2$ & 10.630 &  \\
                 & $4\, ^3\!P_2$ & 10.812 &  \\ 
                 & $3\, ^3\!F_2$ & 10.862 &  \\ 
\hline
$J^P=2^-$\\
 \hline
  $\Upsilon_2(1D)$ & $1\, ^3\!D_2$ & 10.171 & 10.1637 \\
                  & $1\, ^1\!D_2$ & 10.172 &  \\
                  & $2\, ^3\!D_2$ & 10.468 &  \\
                  & $2\, ^1\!D_2$ & 10.469 &  \\
                  & $3\, ^3\!D_2$ & 10.721 &  \\
                  & $3\, ^1\!D_2$ & 10.722 &  \\
                  & $4\, ^3\!D_2$ & 10.947 &  \\
                  & $4\, ^1\!D_2$ & 10.948 &  \\
 \hline
$J^P=3^-$\\
 \hline
                   & $1\, ^3\!D_3$ & 10.180 &  \\
                  & $2\, ^3\!D_3$ & 10.476 &  \\
                  & $1\, ^3\!G_3$ & 10.542 &  \\
                  & $3\, ^3\!D_3$ & 10.728 &  \\
                  & $2\, ^3\!G_3$ & 10.779 &  \\
                  & $4\, ^3\!D_3$ & 10.954 &  \\ 
                  & $3\, ^3\!G_3$ & 10.995 & \\
\hline
$J^P=3^+$\\
 \hline
                  & $1\, ^3\!F_3$ & 10.374 &  \\
                  & $1\, ^1\!F_3$ & 10.375 &  \\
                  & $2\, ^3\!F_3$ & 10.635 &  \\
                  & $2\, ^1\!F_3$ & 10.635 &  \\
                  & $3\, ^3\!F_3$ & 10.866 &  \\
                  & $3\, ^1\!F_3$ & 10.866 &  \\
\end{tabular} 
\end{ruledtabular}
\end{minipage}
\end{table*}

\begin{figure*}[htb]
\centering
\centering
\includegraphics[width=\linewidth]{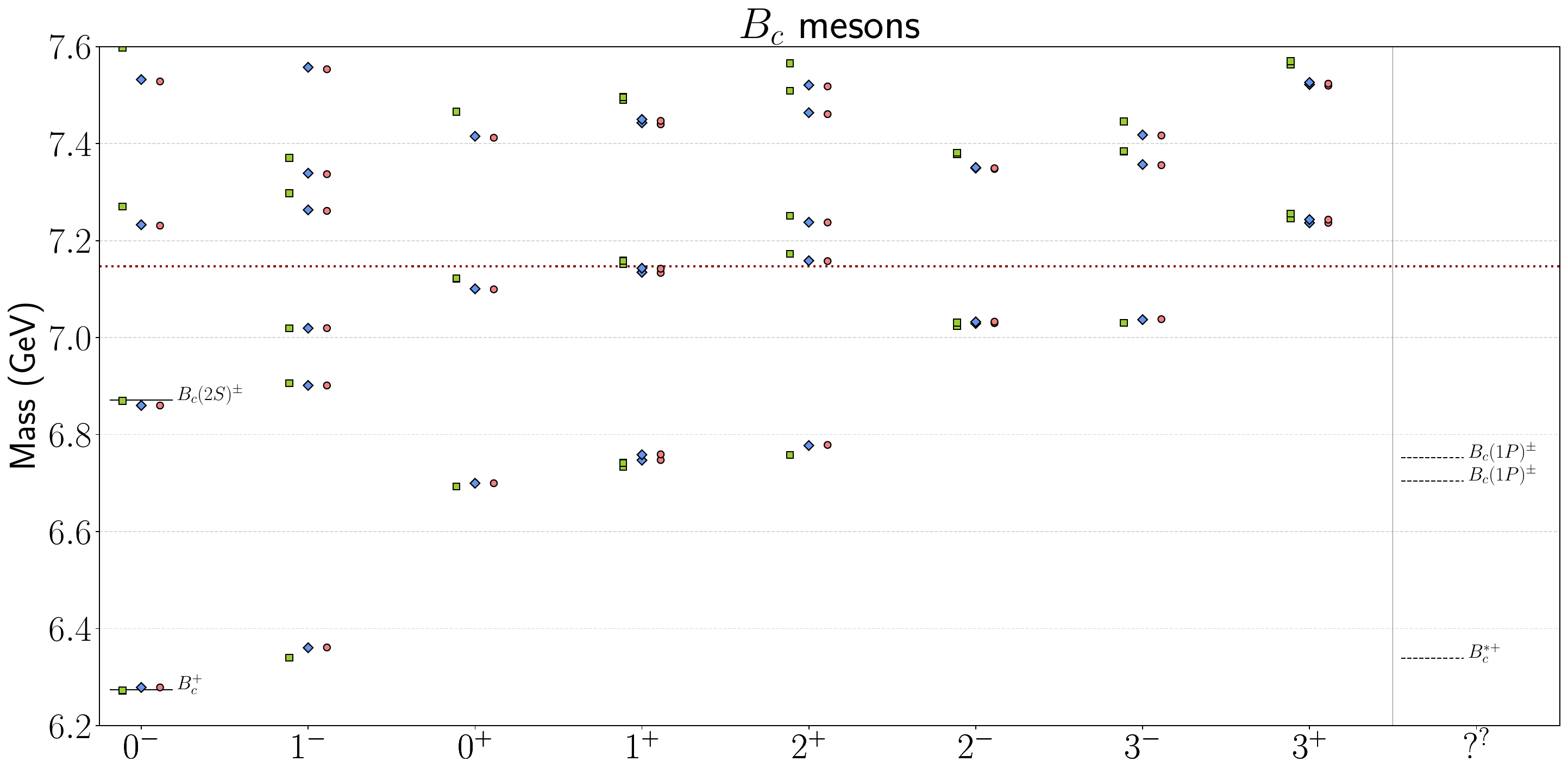}
	\caption{(Color online) Masses of $B_c$ mesons with $J^P$ from $0^\pm$ to $3^\pm$. Measured states with unknown quantum numbers are shown in an extra column labeled ``$?^?$''. The results of fits to pseudoscalar (green squares), non-axial (blue diamonds), and to states of all $J^P$ (red circles), are compared with established (solid lines) and unconfirmed (dashed lines) experimental masses, with the grey shading displaying the width. The red dotted line indicates the lowest open-flavor threshold.}
\label{fig:bc}
\end{figure*}

\begin{table*}[ht]
\caption{$B_c$ mesons masses. All states are coupled, and the dominant state with quantum numbers $n\, ^{2S+1}\!L_J$ is the one with the highest probability. }
\label{tab:bc}
\begin{minipage}[t]{0.45\textwidth}
\vspace{0pt}
\begin{ruledtabular}
 \begin{tabular}{r|ccc}
  & Dominant & & \\
Meson  & state & This work & PDG  \\
\hline
$J^P=0^-$\\
\hline
$B_c$            & $1\, ^1\!S_0$  & 6.279 & 6.27447 \\
$B_c(2S)$     & $2\, ^1\!S_0$  & 6.860 & 6.8712 \\
              & $3\, ^1\!S_0$  & 7.231 &  \\
              & $4\, ^1\!S_0$  & 7.528 &  \\
              & $5\, ^1\!S_0$  & 7.786 &  \\
              & $6\, ^1\!S_0$  & 8.018 &  \\
  \hline
 $J^P=1^-$\\
\hline
                & $1\, ^3\!S_1$ & 6.361 &          \\
                & $2\, ^3\!S_1$ & 6.901 &          \\
                & $1\, ^3\!D_1$ & 7.020 &          \\
                & $3\, ^3\!S_1$ & 7.261 &          \\
                & $2\, ^3\!D_1$ & 7.337 &          \\
                & $4\, ^3\!S_1$ & 7.553 &          \\
                & $3\, ^3\!D_1$ & 7.609 &          \\
                & $5\, ^3\!S_1$ & 7.808 &          \\
                & $4\, ^3\!D_1$ & 7.851 &          \\
                & $6\, ^3\!S_1$ & 8.037 &          \\
 \hline
$J^P=0^+$\\
 \hline
               & $1\, ^3\!P_0$ & 6.700 &  \\
               & $2\, ^3\!P_0$ & 7.100 &  \\
               & $3\, ^3\!P_0$ & 7.412 &  \\
               & $4\, ^3\!P_0$ & 7.680 &  \\
               & $5\, ^3\!P_0$ & 7.919 &  \\
\hline
$J^P=1^+$\\
 \hline
             & $1\, ^3\!P_1$ & 6.748 &  \\
             & $1\, ^1\!P_1$ & 6.759 &  \\
             & $2\, ^3\!P_1$ & 7.134 &  \\
             & $2\, ^1\!P_1$ & 7.142 &  \\
             & $3\, ^3\!P_1$ & 7.440 &  \\
             & $3\, ^1\!P_1$ & 7.447 &  \\
             & $4\, ^3\!P_1$ & 7.704 &  \\
             & $4\, ^1\!P_1$ & 7.710 &  \\
\end{tabular} 
\end{ruledtabular}
\end{minipage}
\hfill
\begin{minipage}[t]{0.45\textwidth}
\vspace{0pt}
\begin{ruledtabular}
 \begin{tabular}{r|ccc}
  & Dominant & & \\
Meson  & state & This work & PDG  \\
\hline
 $J^P=2^+$\\
 \hline
                 & $1\, ^3\!P_2$ & 6.779 & \\
                 & $2\, ^3\!P_2$ & 7.158 &  \\
                 & $1\, ^3\!F_2$ & 7.237 &  \\
                 & $3\, ^3\!P_2$ & 7.461 & \\
                 & $2\, ^3\!F_2$ & 7.518 &  \\
                 & $4\, ^3\!P_2$ & 7.723 &  \\ 
                 & $3\, ^3\!F_2$ & 7.767 &  \\ 
                 & $5\, ^3\!P_2$ & 7.958 &  \\ 
                 & $4\, ^3\!F_2$ & 7.994 &  \\ 
\hline
$J^P=2^-$\\
 \hline
                  & $1\, ^3\!D_2$ & 7.030 &  \\
                  & $1\, ^1\!D_2$ & 7.033 &  \\
                  & $2\, ^3\!D_2$ & 7.348 &  \\
                  & $2\, ^1\!D_2$ & 7.349 &  \\
                  & $3\, ^3\!D_2$ & 7.619 &  \\
                  & $3\, ^1\!D_2$ & 7.620 &  \\
                  & $4\, ^3\!D_2$ & 7.861 &  \\
                  & $4\, ^1\!D_2$ & 7.863 &  \\
 \hline
$J^P=3^-$\\
 \hline
                  & $1\, ^3\!D_3$ & 7.038 &  \\
                  & $2\, ^3\!D_3$ & 7.356 &  \\
                  & $1\, ^3\!G_3$ &  7.417 &  \\
                  & $3\, ^3\!D_3$ & 7.627 &  \\
                  & $2\, ^3\!G_3$ &  7.674 &  \\
                  & $4\, ^3\!D_3$ & 7.869 &  \\ 
                  & $3\, ^3\!G_3$ &  7.907 & \\
\hline
$J^P=3^+$\\
 \hline
                  & $1\, ^3\!F_3$ & 7.237 &  \\
                  & $1\, ^1\!F_3$ & 7.243 &  \\
                  & $2\, ^3\!F_3$ & 7.520 &  \\
                  & $2\, ^1\!F_3$ & 7.524 &  \\
                  & $3\, ^3\!F_3$ & 7.770 &  \\
                  & $3\, ^1\!F_3$ & 7.773 &  \\
                  & $4\, ^3\!F_3$ & 7.997 &  \\
                  & $4\, ^1\!F_3$ & 7.999 &  \\
                    \hline
                      \hline
                   & Possible & & \\
                    $J^P=?^?$& assignment(s)&  This work &  Exp.\\
                    \hline
                    $B_c^{*+}$ & &  &6.3390  \\
                     & $1\, ^3\!S_1$ & 6.361 &          \\

                  \hline
                  $B_c(1P)^\pm$ & &  &6.7048  \\& $1\, ^3\!P_0\, (0^+)$ & 6.700 &\\
                  & $1\, ^3\!P_1 \, (1^+)$ & 6.748 &  \\
                   $B_c(1P)^\pm$ & &  &6.7524  \\
                    
                    & $1\, ^1\!P_1 \, (1^+)$ & 6.759 &  \\
                      & $1\, ^3\!P_2 \, (2^+)$ & 6.779 & 
\end{tabular} 
\end{ruledtabular}
\end{minipage}
\end{table*}

\begin{figure*}[htb]
\centering
\centering
\includegraphics[width=\linewidth]{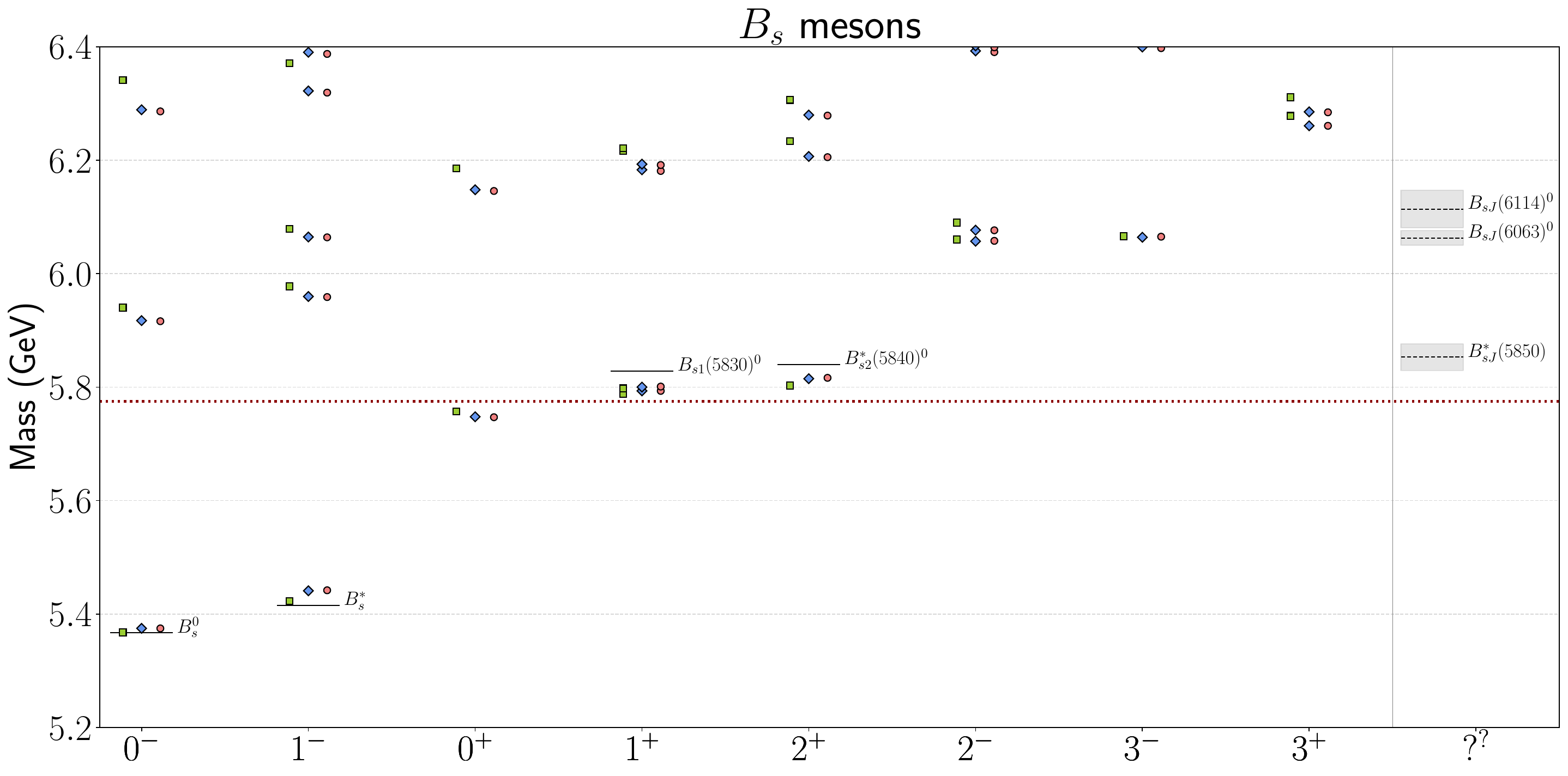}
	\caption{(Color online) Masses of $B_s$ mesons with $J^P$ from $0^\pm$ to $3^\pm$. Measured states with unknown quantum numbers are shown in an extra column labeled ``$?^?$''. The results of fits to pseudoscalar (green squares), non-axial (blue diamonds), and to states of all $J^P$ (red circles), are compared with established (solid lines) and unconfirmed (dashed lines) experimental masses, with the grey shading displaying the width. The red dotted line indicates the lowest open-flavor threshold.}
\label{fig:bs}
\end{figure*}

\begin{table*}[t]
\caption{$B_s$ meson masses. All states are coupled, and the dominant state with quantum numbers $n\, ^{2S+1}\!L_J$ is the one with the highest probability. }
\label{tab:bs}
\begin{minipage}[t]{0.45\textwidth}
\vspace{0pt}
\begin{ruledtabular}
 \begin{tabular}{r|ccc}
  & Dominant & & \\
Meson  & state & This work & PDG  \\
\hline
$J^P=0^-$\\
\hline
$B_s$         & $1\, ^1\!S_0$  & 5.375 & 5.36691 \\
              & $2\, ^1\!S_0$  & 5.916 & \\
              & $3\, ^1\!S_0$  & 6.286 &  \\
              & $4\, ^1\!S_0$  & 6.586 &  \\
              & $5\, ^1\!S_0$  & 6.846 &  \\
              & $6\, ^1\!S_0$  & 7.081 &  \\
  \hline
 $J^P=1^-$\\
\hline
$B_s^*$         & $1\, ^3\!S_1$ & 5.442 & 5.4154         \\
                & $2\, ^3\!S_1$ & 5.959 &          \\
                & $1\, ^3\!D_1$ & 6.064 &          \\
                & $3\, ^3\!S_1$ & 6.319 &          \\
                & $2\, ^3\!D_1$ & 6.387 &          \\
                & $4\, ^3\!S_1$ & 6.614 &          \\
                & $3\, ^3\!D_1$ & 6.664 &          \\
                & $5\, ^3\!S_1$ & 6.871 &          \\
                & $4\, ^3\!D_1$ & 6.910 &          \\
                & $6\, ^3\!S_1$ & 7.103 &          \\
 \hline
$J^P=0^+$\\
 \hline
               & $1\, ^3\!P_0$ & 5.747 & \\
               & $2\, ^3\!P_0$ & 6.146 &  \\
               & $3\, ^3\!P_0$ & 6.463 &  \\
               & $4\, ^3\!P_0$ & 6.735 &  \\
               & $5\, ^3\!P_0$ & 6.977 &  \\
\hline
$J^P=1^+$\\
 \hline
$B_{s1}(5830)$    & $1\, ^3\!P_1$ & 5.793 & 5.82865 \\
             & $1\, ^1\!P_1$ & 5.801 &  \\
             & $2\, ^3\!P_1$ & 6.181 &  \\
             & $2\, ^1\!P_1$ & 6.192 &  \\
             & $3\, ^3\!P_1$ & 6.492 &  \\
             & $3\, ^1\!P_1$ & 6.502 &  \\
             & $4\, ^3\!P_1$ & 6.760 &  \\
             & $4\, ^1\!P_1$ & 6.770 &  \\
\end{tabular} 
\end{ruledtabular}
\end{minipage}
\hfill
\begin{minipage}[t]{0.45\textwidth}
\vspace{0pt}
\begin{ruledtabular}
 \begin{tabular}{r|ccc}
  & Dominant & & \\
Meson  & state & This work & PDG  \\
\hline
 $J^P=2^+$\\
 \hline
$B_{s2}^*(5840)$ & $1\, ^3\!P_2$ & 5.816 & 5.83988 \\
                 & $2\, ^3\!P_2$ & 6.205 &  \\
                 & $1\, ^3\!F_2$ & 6.279 &  \\
                 & $3\, ^3\!P_2$ & 6.515 & \\
                 & $2\, ^3\!F_2$ & 6.567 &  \\
                 & $4\, ^3\!P_2$ & 6.782 &  \\ 
                 & $3\, ^3\!F_2$ & 6.822 &  \\ 
\hline
$J^P=2^-$\\
 \hline
                  & $1\, ^1\!D_2- 1\, ^3\!D_2$ & 6.058 &  \\
                  & $1\, ^3\!D_2-1\, ^1\!D_2$ & 6.077 &  \\
                  & $2\, ^1\!D_2-2\, ^3\!D_2$ & 6.391 &  \\
                  & $2\, ^3\!D_2-2\, ^1\!D_2$ & 6.399 &  \\
                  & $3\, ^1\!D_2- 3\, ^3\!D_2$ & 6.671 &  \\
                  & $3\, ^3\!D_2-3\, ^1\!D_2$ & 6.675 &  \\
                  & $4\, ^3\!D_2$ & 6.919 &  \\
                  & $4\, ^1\!D_2$ & 6.921 &  \\
 \hline
$J^P=3^-$\\
 \hline
                  & $1\, ^3\!D_3$ & 6.065 &  \\
                  & $2\, ^3\!D_3$ & 6.398 &  \\
                  & $1\, ^3\!G_3$ & 6.456 &  \\
                  & $3\, ^3\!D_3$ & 6.678 &  \\
                  & $2\, ^3\!G_3$ & 6.722 &  \\
                  & $4\, ^3\!D_3$ & 6.926 &  \\ 
                  & $3\, ^3\!G_3$ & 6.961 &  \\
                  & $5\, ^3\!D_3$ & 7.151 &  \\ 
\hline
$J^P=3^+$\\
 \hline
                  & $1\, ^1\!F_3- 1\, ^3\!F_3$ & 6.261 &  \\
                  & $1\, ^3\!F_3- 1\, ^1\!F_3$ & 6.284 &  \\
                  & $2\, ^1\!F_3- 2\, ^3\!F_3$ & 6.558 &  \\
                  & $2\, ^3\!F_3- 2\, ^1\!F_3$ & 6.573 &  \\
                  & $3\, ^1\!F_3- 3\, ^3\!F_3$ & 6.818 &  \\
                  & $3\, ^3\!F_3- 3\, ^1\!F_3$ & 6.828 &  \\
                  & $4\, ^1\!F_3- 4\, ^3\!F_3$ & 7.052 &  \\
                  & $4\, ^3\!F_3- 4\, ^1\!F_3$ & 7.059 &  \\
                  \hline
                  \hline
                  & Possible & & \\
                  $J^P=?^?$& assignment(s)&  This work &  PDG\\
                  \hline
                 $B_{sJ}^* (5850)$ &$1\, ^1\!P_1 \, (1^+)$ & 5.801 & 5.83992  \\

                  \hline
                  
                  $B_{sJ}(6063)^0$ 
                 & &  &6.0635  \\
                 $B_{sJ}(6114)^0$ & & & 6.114   \\
                 & $1\, ^1\!D_2- 1\, ^3\!D_2 \, (2^-)$ & 6.058 &\\
                  & $1\, ^3\!D_1 \, (1^-)$ & 6.064\\
                  & $1\, ^3\!D_3 \,(3^-)$ & 6.065 &  \\
                 
                   & $1\, ^3\!D_2-1\, ^1\!D_2 \, (2^-)$ & 6.077 &  \\

\end{tabular} 
\end{ruledtabular}
\end{minipage}
\end{table*}

\begin{figure*}[htb]
\centering
\centering
\includegraphics[width=\linewidth]{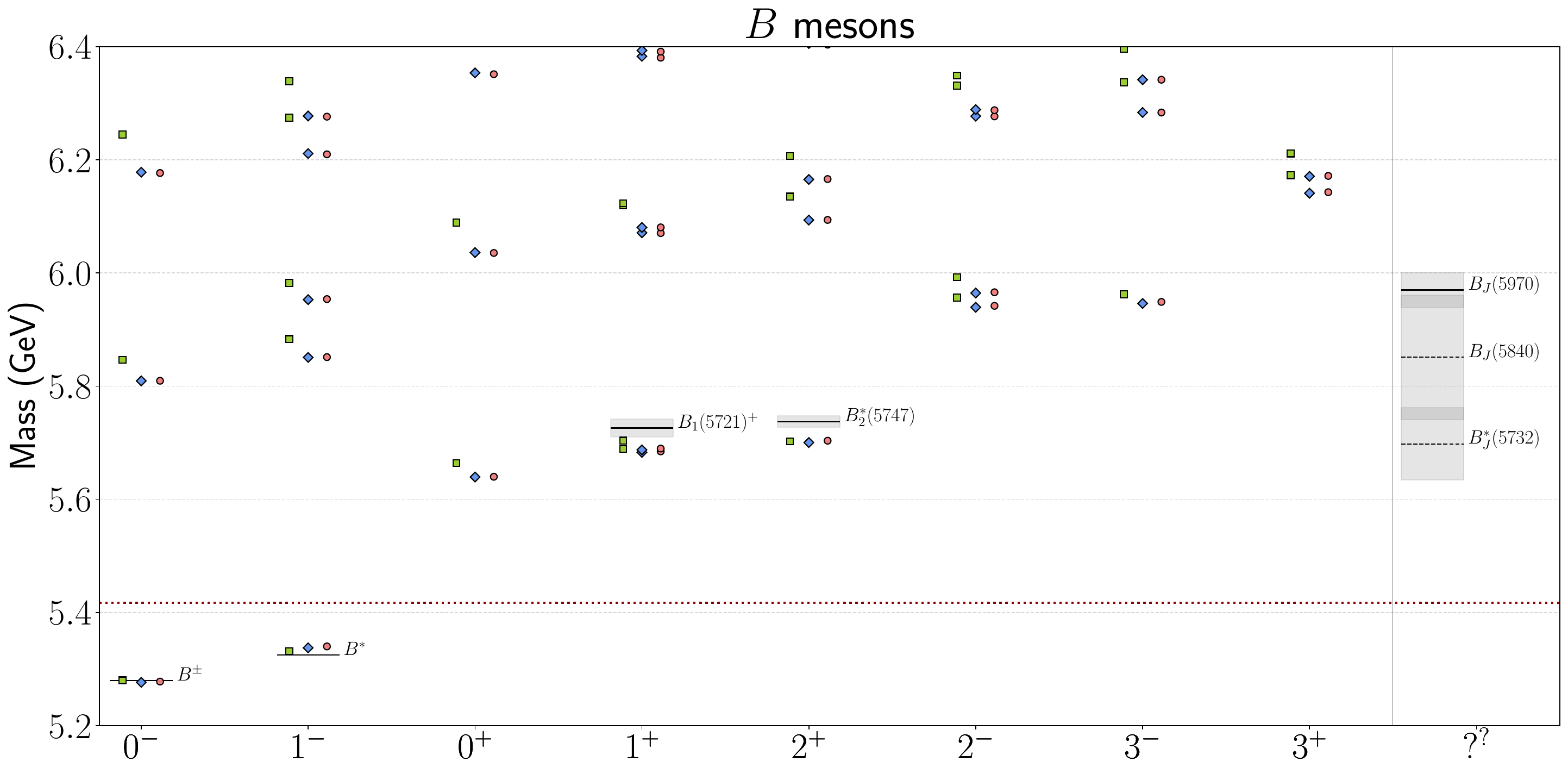}
	\caption{(Color online) Masses $B$ mesons with $J^P$ from $0^\pm$ to $3^\pm$. Measured states with unknown quantum numbers are shown in an extra column labeled ``$?^?$''. The results of fits to pseudoscalar (green squares), non-axial (blue diamonds), and to states of all $J^P$ (red circles), are compared with established (solid lines) and unconfirmed (dashed lines) experimental masses, with the grey shading displaying the width. The red dotted line indicates the lowest open-flavor threshold.}
\label{fig:bq}
\end{figure*}

\begin{table*}[t]
\caption{$B$ meson masses. All states are coupled, and the dominant state with quantum numbers $n\, ^{2S+1}\!L_J$ is the one with the highest probability. }
\label{tab:bq}
\begin{minipage}[t]{0.45\textwidth}
\vspace{0pt}
\begin{ruledtabular}
 \begin{tabular}{r|ccc}
  & Dominant & & \\
Meson  & state & This work & PDG  \\
\hline
$J^P=0^-$\\
\hline
$B$           & $1\, ^1\!S_0$  & 5.278 & 5.27941 \\
              & $2\, ^1\!S_0$  & 5.810 & \\
              & $3\, ^1\!S_0$  & 6.177 &  \\
              & $4\, ^1\!S_0$  & 6.475 &  \\
              & $5\, ^1\!S_0$  & 6.735 &  \\
              & $6\, ^1\!S_0$  & 6.968 &  \\
  \hline
 $J^P=1^-$\\
\hline
$B^*$           & $1\, ^3\!S_1$ & 5.340 & 5.32475         \\
                & $2\, ^3\!S_1$ & 5.851 &          \\
                & $1\, ^3\!D_1$ & 5.954 &          \\
                & $3\, ^3\!S_1$ & 6.210 &          \\
                & $2\, ^3\!D_1$ & 6.276 &          \\
                & $4\, ^3\!S_1$ & 6.503 &          \\
                & $3\, ^3\!D_1$ & 6.552 &          \\
                & $5\, ^3\!S_1$ & 6.759 &          \\
                & $4\, ^3\!D_1$ & 6.798 &          \\
                & $6\, ^3\!S_1$ & 6.991 &          \\
 \hline
$J^P=0^+$\\
 \hline
               & $1\, ^3\!P_0$ & 5.640 & \\
               & $2\, ^3\!P_0$ & 6.035 &  \\
               & $3\, ^3\!P_0$ & 6.351 &  \\
               & $4\, ^3\!P_0$ & 6.622 &  \\
               & $5\, ^3\!P_0$ & 6.865 &  \\
\hline
$J^P=1^+$\\
 \hline
$B_1(5721)$  & $1\, ^3\!P_1$ & 5.685 & 5.7260 \\
             & $1\, ^1\!P_1$ & 5.690 &  \\
             & $2\, ^3\!P_1$ & 6.070 &  \\
             & $2\, ^1\!P_1$ & 6.081 &  \\
             & $3\, ^3\!P_1$ & 6.380 &  \\
             & $3\, ^1\!P_1$ & 6.391 &  \\
             & $4\, ^3\!P_1$ & 6.648 &  \\
             & $4\, ^1\!P_1$ & 6.658 &  \\
\end{tabular} 
\end{ruledtabular}
\end{minipage}
\hfill
\begin{minipage}[t]{0.45\textwidth}
\vspace{0pt}
\begin{ruledtabular}
 \begin{tabular}{r|ccc}
  & Dominant & & \\
Meson  & state & This work & PDG  \\
\hline
 $J^P=2^+$\\
 \hline
$B_2^*(5747)$    & $1\, ^3\!P_2$ & 5.704 & 5.7373 \\
                 & $2\, ^3\!P_2$ & 6.094 &  \\
                 & $1\, ^3\!F_2$ & 6.166 &  \\
                 & $3\, ^3\!P_2$ & 6.404 & \\
                 & $2\, ^3\!F_2$ & 6.455 &  \\
                 & $4\, ^3\!P_2$ & 6.670 &  \\ 
                 & $3\, ^3\!F_2$ & 6.709 &  \\ 
\hline
$J^P=2^-$\\
 \hline
                  & $1\, ^1\!D_2- 1\, ^3\!D_2$ & 5.942 &  \\
                  & $1\, ^3\!D_2-1\, ^1\!D_2$  & 5.966 &  \\
                  & $2\, ^1\!D_2- 2\, ^3\!D_2$ & 6.277 &  \\
                  & $2\, ^3\!D_2-2\, ^1\!D_2$  & 6.288 &  \\
                  & $3\, ^1\!D_2- 3\, ^3\!D_2$ & 6.558 &  \\
                  & $3\, ^3\!D_2-3\, ^1\!D_2$  & 6.563 &  \\
                  & $4\, ^3\!D_2$            & 6.806 &  \\
                  & $4\, ^1\!D_2$            & 6.808 &  \\
 \hline
$J^P=3^-$\\
 \hline
                  & $1\, ^3\!D_3$ & 5.949 &  \\
                  & $2\, ^3\!D_3$ & 6.284 &  \\
                  & $1\, ^3\!G_3$ & 6.342 &  \\
                  & $3\, ^3\!D_3$ & 6.564 &  \\
                  & $2\, ^3\!G_3$ & 6.608 &  \\
                  & $4\, ^3\!D_3$ & 6.812 &  \\ 
                  & $3\, ^3\!G_3$ & 6.847 &  \\
                  & $5\, ^3\!D_3$ & 7.038 &  \\ 
\hline
$J^P=3^+$\\
 \hline
                   & $1\, ^1\!F_3- 1\, ^3\!F_3$ & 6.143 &  \\
                  & $1\, ^3\!F_3- 1\, ^1\!F_3$ & 6.172 &  \\
                  & $2\, ^1\!F_3- 2\, ^3\!F_3$ & 6.443 &  \\
                  & $2\, ^3\!F_3- 2\, ^1\!F_3$ & 6.460 &  \\
                  & $3\, ^1\!F_3- 3\, ^3\!F_3$ & 6.703 &  \\
                  & $3\, ^3\!F_3-3 \, ^1\!F_3$ & 6.715 &  \\
                  & $4\, ^1\!F_3-4 \, ^3\!F_3$ & 6.937 &  \\
                  & $4\, ^3\!F_3-4 \, ^1\!F_3$ & 6.946 & \\ \hline
                  \hline
                  & Possible & & \\
                  $J^P=?^?$& assignment(s)&  This work &  PDG\\
                  \hline
                  $B_{J}^* (5732)$ & & & 5.698 
                   \\
                  & $1\, ^3\!P_0 \, (0^+)$ & 5.640 & \\
                   & $1\, ^3\!P_1 \, (1^+)$ & 5.685 &\\
                  & $1\, ^1\!P_1 \, (1^+)$ & 5.690 &  \\
                   & $1\, ^3\!P_2 \, (2^+)$ & 5.704 & \\
                   
                  \hline
                  
                  $B_{J}(5840)$ 
                  & &  &5.851  \\
                   & $2\, ^1\!S_0\, (0^-)$  & 5.810 &\\
                    & $2\, ^3\!S_1 \, (1^-)$ & 5.851 &          \\
                      \hline
                    $B_{J}(5970)$ 
                    & &  &5.965  \\
                     & $1\, ^1\!D_2- 1\, ^3\!D_2 \,(2^-)$ & 5.942 &  \\
                     & $1\, ^3\!D_3  \,(3^-)$ & 5.949 &  \\& $1\, ^3\!D_1 \, (1^-)$ & 5.954 &          \\
                    & $1\, ^3\!D_2-1\, ^1\!D_2 \,(2^-)$  & 5.966 &

\end{tabular} 
\end{ruledtabular}
\end{minipage}
\end{table*}

\begin{figure*}[htb]
\centering
\centering
\includegraphics[width=\linewidth]{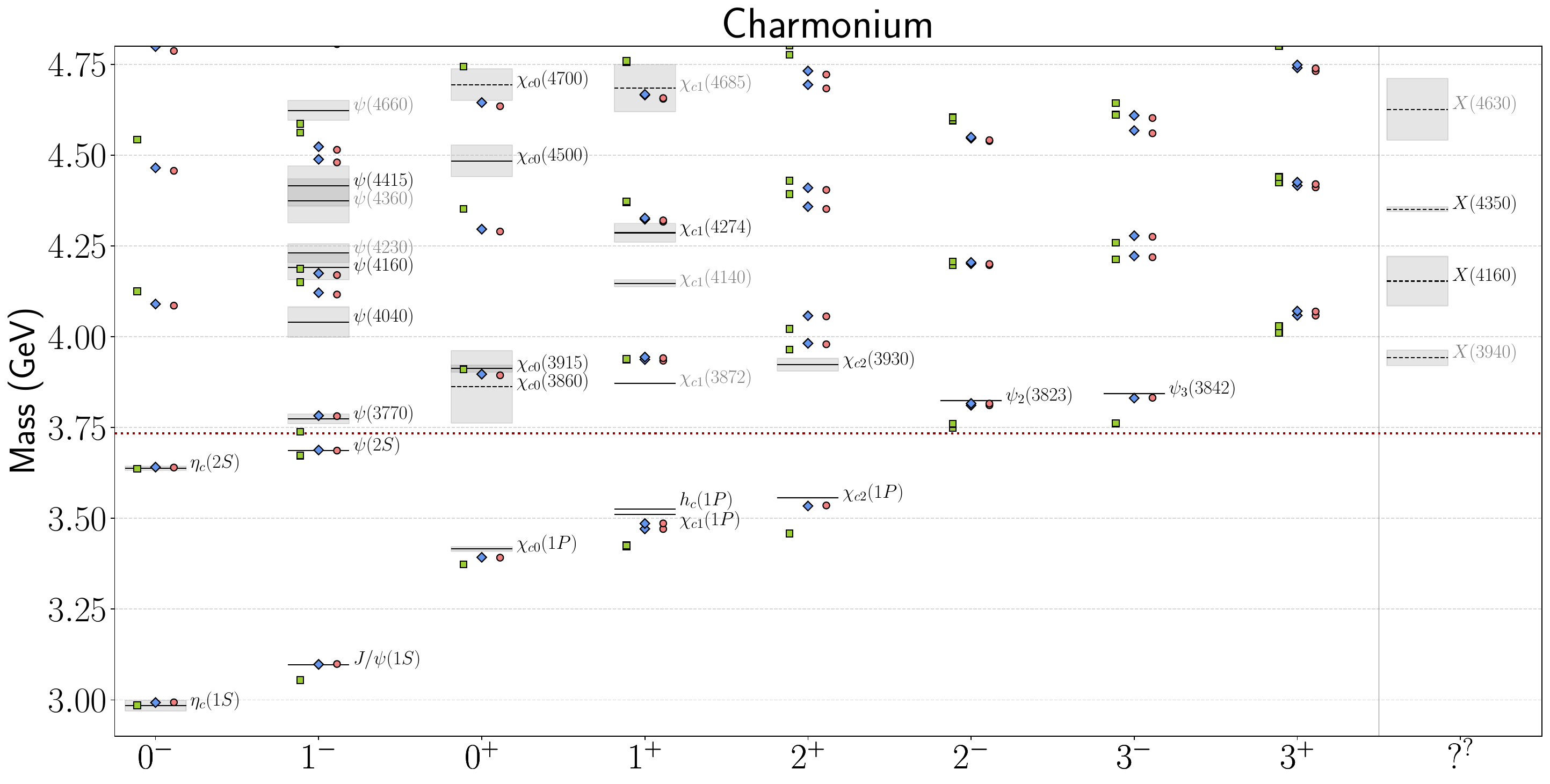}
	\caption{(Color online) Masses of charmonium with $J^P$ from $0^\pm$ to $3^\pm$. Measured states with unknown quantum numbers are shown in an extra column labeled ``$?^?$''. The results of fits to pseudoscalar (green squares), non-axial (blue diamonds), and to states of all $J^P$ (red circles), are compared with established (solid lines) and unconfirmed (dashed lines) experimental masses, with the grey shading displaying the width. The red dotted line indicates the lowest open-charm threshold. State names in grey font indicate possible non-$q\bar q$ (exotic) candidates.}
\label{fig:cc}
\end{figure*}

\begin{table*}[t]
\caption{Charmonium masses. All states are coupled, and the dominant state with quantum numbers $n\, ^{2S+1}\!L_J$ is the one with the highest probability. }
\label{tab:cc}
\begin{minipage}[t]{0.45\textwidth}
\vspace{0pt}
\begin{ruledtabular}
 \begin{tabular}{r|ccc}
  & Dominant & & \\
Meson  & state & This work & PDG  \\
\hline
$J^P=0^-$\\
\hline
$\eta_c(1S)$  & $1\, ^1\!S_0$  & 2.993 & 2.98409 \\
$\eta_c(2S)$  & $2\, ^1\!S_0$  & 3.640 & 3.6378 \\
              & $3\, ^1\!S_0$  & 4.085 &  \\
              & $4\, ^1\!S_0$  & 4.457 &  \\
              & $5\, ^1\!S_0$  & 4.787 &  \\
              & $6\, ^1\!S_0$  & 5.090 &  \\
 \hline
 $J^P=1^-$\\
\hline
$J/\psi(1S)$    & $1\, ^3\!S_1$ & 3.099 & 3.096900  \\
$\psi(2S)$      & $2\, ^3\!S_1$ & 3.686 & 3.686097 \\
$\psi(3770)$    & $1\, ^3\!D_1$ & 3.781 & 3.7737    \\
$\psi(4040)$    &                       &           & 4.040 \\
                & $3\, ^3\!S_1$ & 4.116 &          \\
$\psi(4160)$   &                        &           & 4.191 \\
                & $2\, ^3\!D_1$ & 4.170 &          \\
$\psi(4230)$   &                        &           & 4.2217 \\
$\psi(4360)$   &                        &           & 4.373 \\
$\psi(4415)$   &                        &           & 4.415 \\
                & $4\, ^3\!S_1$ & 4.480 &          \\
                & $3\, ^3\!D_1$ & 4.515 &          \\
$\psi(4660)$   &                        &           & 4.623 \\
                & $5\, ^3\!S_1$ & 4.805 &          \\
                & $4\, ^3\!D_1$ & 4.830 &          \\
                & $6\, ^3\!S_1$ & 5.105 &          \\
 \hline
$J^P=0^+$\\
 \hline
$\chi_{c0}(1P)$ & $1\, ^3\!P_0$ & 3.392 & 3.41550 \\
$\chi_{c0}(3860)$ &  &  &  3.862 \\
$\chi_{c0}(3915)$ & $2\, ^3\!P_0$ & 3.894 & 3.91221 \\
               & $3\, ^3\!P_0$ & 4.289 &  \\
$\chi_{c0}(4500)$ & & & 4.484 \\
               & $4\, ^3\!P_0$ & 4.634 &  \\
$\chi_{c0}(4700)$ & & & 4.708 \\     
               & $5\, ^3\!P_0$ & 4.948 &  \\
\hline
$J^P=1^+$\\
 \hline
$\chi_{c1}(1P)$ & $1\, ^3\!P_1$ & 3.470 & 3.51067 \\
$h_c(1P)$       & $1\, ^1\!P_1$ & 3.486 & 3.52537 \\
$\chi_{c1}(3872)$ &   &   & 3.87164 \\
  & $2\, ^3\!P_1$ & 3.934 &   \\
             & $2\, ^1\!P_1$ & 3.941 &  \\
$\chi_{c1}(4140)$ & & & 4.1465 \\
$\chi_{c1}(4274)$ & & & 4.290 \\
             & $3\, ^3\!P_1$ & 4.317 &  \\
             & $3\, ^1\!P_1$ & 4.320 &  \\
             & $4\, ^3\!P_1$ & 4.655 &  \\
             & $4\, ^1\!P_1$ & 4.657 &  \\
$\chi_{c1}(4685)$ & & & 4.677 \\
\end{tabular} 
\end{ruledtabular}
\end{minipage}
\hfill
\begin{minipage}[t]{0.45\textwidth}
\vspace{0pt}
\begin{ruledtabular}
 \begin{tabular}{r|ccc}
  & Dominant & & \\
Meson  & state & This work & PDG  \\
\hline
 $J^P=2^+$\\
 \hline
$\chi_{c2}(1P)$  & $1\, ^3\!P_2$ & 3.535 & 3.55617 \\
$\chi_{c2}(3930)$  & $2\, ^3\!P_2$ & 3.979 & 3.9225 \\
                 & $1\, ^3\!F_2$ & 4.056 &  \\
                 & $3\, ^3\!P_2$ & 4.352 & \\
                 & $2\, ^3\!F_2$ & 4.404 &  \\
                 & $4\, ^3\!P_2$ & 4.683 &  \\ 
                 & $3\, ^3\!F_2$ & 4.722 &  \\ 
\hline
$J^P=2^-$\\
 \hline
 $\psi_2(3823)$  & $1\, ^3\!D_2-1\, ^1\!D_2$ & 3.811 & 3.82351 \\
                  & $1\, ^1\!D_2- 1\, ^3\!D_2$ & 3.816 &  \\
                  & $2\, ^3\!D_2-2\, ^1\!D_2$ & 4.197 &  \\
                  & $2\, ^1\!D_2-2 \, ^3\!D_2$ & 4.200 &  \\
                  & $3\, ^3\!D_2-3\, ^1\!D_2$ & 4.538 &  \\
                  & $3\, ^1\!D_2-3 \, ^3\!D_2$ & 4.541 &  \\
                  & $4\, ^3\!D_2-4\, ^1\!D_2$ & 4.851 &  \\
                  & $4\, ^1\!D_2- 4\, ^3\!D_2$ & 4.853 &  \\
 \hline
$J^P=3^-$\\
 \hline
$\psi_3(3842)$    & $1\, ^3\!D_3$ & 3.832 & 3.84271 \\
                  & $2\, ^3\!D_3$ & 4.219 &  \\
                  & $1\, ^3\!G_3$ & 4.275 &  \\
                  & $3\, ^3\!D_3$ & 4.560 &  \\
                  & $2\, ^3\!G_3$ & 4.602 &  \\
                  & $4\, ^3\!D_3$ & 4.871 &  \\ 
                  & $3\, ^3\!G_3$ & 4.904 & \\
\hline
$J^P=3^+$\\
 \hline
                  & $1\, ^1\!F_3- 1\, ^3\!F_3$ & 4.058 &  \\
                  & $1\, ^3\!F_3- 1\, ^1\!F_3$ & 4.069 &  \\
                  & $2\, ^1\!F_3- 2\, ^3\!F_3$ & 4.411 &  \\
                  & $2\, ^3\!F_3- 2\, ^1\!F_3$ & 4.420 &  \\
                  & $3\, ^1\!F_3- 3\, ^3\!F_3$ & 4.731 &  \\
                  & $3\, ^3\!F_3- 3\, ^1\!F_3$ & 4.739 &  \\
                  & $4\, ^1\!F_3- 4\, ^3\!F_3$ & 5.028 &  \\
                  & $4\, ^3\!F_3- 4\, ^1\!F_3$ & 5.034 &  \\
                  \hline
                  \hline
                  & Possible & & \\
                  $J^P=?^?$& assignment(s)&  This work &  PDG\\
                  \hline
                  $X(3940)$ &
                 & &3.942\\
                  & $2\, ^3\!P_1 \, (1^{+})$ & 3.934 &  \\
                 &  $3\, ^1\!S_0\, (0^-)$  & 4.085 &\\
                  \hline
                  $X(4160)$ &
                  & &4.153\\
                  & $2\, ^3\!D_2-2\, ^1\!D_2\, (2^-)$  & 4.197 &  \\
                  & $2\, ^1\!D_2-2 \, ^3\!D_2\, (2^-)$ & 4.200 &  \\
                   \hline
                  $X(4350)$ &
                  & &4.3506\\
                   & $3\, ^3\!P_0 \, (0^+)$ & 4.289 &  \\
                   & $3\, ^3\!P_2 \, (2^+)$ & 4.352 & \\
                   & $2\, ^3\!F_2\, (2^+)$ & 4.404 &  \\
                  \hline
                 $X(4630)$ &
                 & &4.636\\
                 & $3\, ^3\!D_2-3\, ^1\!D_2\, (2^-)$ & 4.538 &  \\
                 & $3\, ^1\!D_2-3 \, ^3\!D_2\, (2^-)$ & 4.541 &

\end{tabular} 
\end{ruledtabular}
\end{minipage}
\end{table*}

\begin{figure*}[htb]
\centering
\centering
\includegraphics[width=\linewidth]{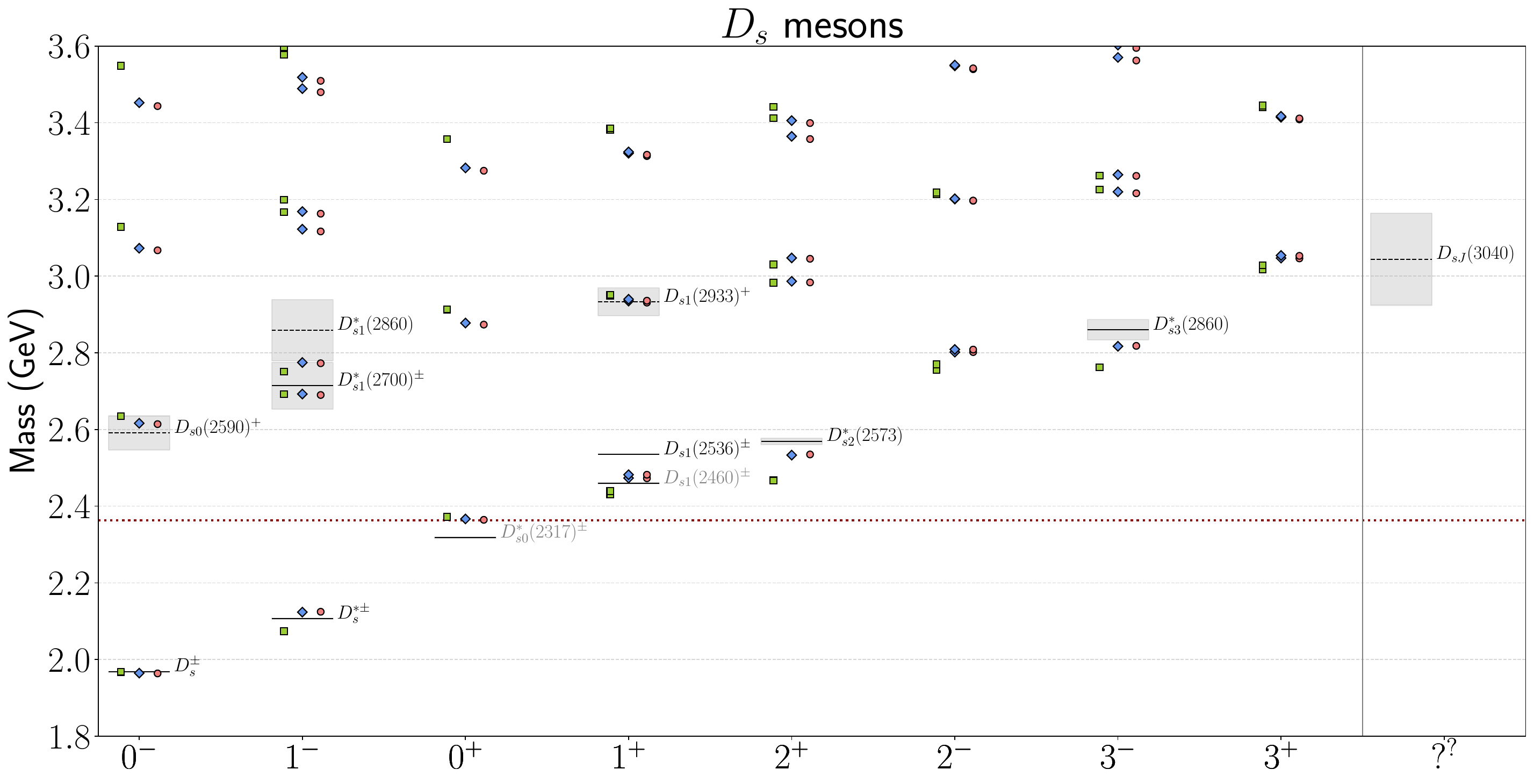}
	\caption{(Color online) Masses of $D_s$ mesons with $J^P$ from $0^\pm$ to $3^\pm$. Measured states with unknown quantum numbers are shown in an extra column labeled ``$?^?$''. The results of fits to pseudoscalar (green squares), non-axial (blue diamonds), and to states of all $J^P$ (red circles), are compared with established (solid lines) and unconfirmed (dashed lines) experimental masses, with the grey shading displaying the width. The red dotted line indicates the lowest open-flavor threshold. State names in grey font indicate possible non-$q\bar q$ (exotic) candidates.}
\label{fig:cs}
\end{figure*}

\begin{table*}[t]
\caption{$D_s$ meson masses. All states are coupled, and the dominant state with quantum numbers $n\, ^{2S+1}\!L_J$ is the one with the highest probability. }
\label{tab:cs}
\begin{minipage}[t]{0.45\textwidth}
\vspace{0pt}
\begin{ruledtabular}
 \begin{tabular}{r|ccc}
  & Dominant & & \\
Meson  & state & This work & PDG  \\
\hline
$J^P=0^-$\\
\hline
$D_s$          & $1\, ^1\!S_0$  & 1.964 & 1.96835 \\
$D_{s0}(2590)^+$  & $2\, ^1\!S_0$  & 2.614 & 2.591 \\
               & $3\, ^1\!S_0$  & 3.067 &  \\
               & $4\, ^1\!S_0$  & 3.443 &  \\
               & $5\, ^1\!S_0$  & 3.777 &  \\
               & $6\, ^1\!S_0$  & 4.082 &  \\
  \hline
 $J^P=1^-$\\
\hline
$D_s^{*\pm}$         & $1\, ^3\!S_1$ & 2.125 & 2.1066  \\
$D_{s1}^*(2700)^\pm$ & $2\, ^3\!S_1$ & 2.690 & 2.714 \\
$D_{s1}^*(2860)$  & $1\, ^3\!D_1$ & 2.773 &  2.859   \\
                & $3\, ^3\!S_1$ & 3.117 &          \\
                & $2\, ^3\!D_1$ & 3.163 &          \\
                & $4\, ^3\!S_1$ & 3.480 &          \\
                & $3\, ^3\!D_1$ & 3.509 &          \\
                & $5\, ^3\!S_1$ & 3.805 &          \\
                & $4\, ^3\!D_1$ & 3.826 &          \\
                & $6\, ^3\!S_1$ & 4.105 &          \\
 \hline
 \hline
$J^P=0^+$\\
 \hline
 $D_{s0}^*(2317)^\pm$ & & & 2.3178 \\
               & $1\, ^3\!P_0$ & 2.364 & \\
               & $2\, ^3\!P_0$ & 2.874 & \\
               & $3\, ^3\!P_0$ & 3.275 &  \\
               & $4\, ^3\!P_0$ & 3.624 &  \\
               & $5\, ^3\!P_0$ & 3.940 &  \\
\hline
$J^P=1^+$\\
 \hline
$D_{s1}(2460)^\pm$ & & & 2.4595 \\
  & $1\, ^3\!P_1$ & 2.473 &  \\
  & $1\, ^1\!P_1$ & 2.482 & \\
$D_{s1}(2536)^\pm$  &  &  & 2.53511\\
             & $2\, ^3\!P_1$ & 2.931 & \\
$D_{s1} (2933)^+$ &                   & &2.933\\
             & $2\, ^1\!P_1$ & 2.936 &  \\
             & $3\, ^3\!P_1$ & 3.313 &  \\
             & $3\, ^1\!P_1$ & 3.317 &  \\
             & $4\, ^3\!P_1$ & 3.652 &  \\
             & $4\, ^1\!P_1$ & 3.655 &  \\
\end{tabular} 
\end{ruledtabular}
\end{minipage}
\hfill
\begin{minipage}[t]{0.45\textwidth}
\vspace{0pt}
\begin{ruledtabular}
 \begin{tabular}{r|ccc}
  & Dominant & & \\
Meson  & state & This work & PDG  \\
\hline
 $J^P=2^+$\\
 \hline
$D_{s2}^*(2573)$  & $1\, ^3\!P_2$ & 2.535 & 2.5691\\
                 & $2\, ^3\!P_2$ & 2.984 & \\
                 & $1\, ^3\!F_2$ & 3.045 &  \\
                 & $3\, ^3\!P_2$ & 3.358 & \\
                 & $2\, ^3\!F_2$ & 3.399 &  \\
                 & $4\, ^3\!P_2$ & 3.690 &  \\ 
                 & $3\, ^3\!F_2$ & 3.720 &  \\ 
\hline
$J^P=2^-$\\
 \hline
                  & $1\, ^1\!D_2$ & 2.802 & \\
                  & $1\, ^3\!D_2$ & 2.808 &  \\
                  & $2\, ^1\!D_2$ & 3.197 &  \\
                  & $2\, ^3\!D_2$ & 3.197 &  \\
                  & $3\, ^1\!D_2$ & 3.540 &  \\
                  & $3\, ^3\!D_2$ & 3.542 &  \\
                  & $4\, ^1\!D_2$ & 3.853 &  \\
                  & $4\, ^3\!D_2$ & 3.856 &  \\
 \hline
$J^P=3^-$\\
 \hline
                  & $1\, ^3\!D_3$ & 2.818 & \\
                  & $2\, ^3\!D_3$ & 3.216 &  \\
                  & $1\, ^3\!G_3$ & 3.262 &  \\
                  & $3\, ^3\!D_3$ & 3.562 &  \\
                  & $2\, ^3\!G_3$ & 3.595 &  \\
                  & $4\, ^3\!D_3$ & 3.876 &  \\ 
                  & $3\, ^3\!G_3$ & 3.901 & \\
\hline
$J^P=3^+$\\
 \hline
                  & $1\, ^3\!F_3$ & 3.046 &  \\
                  & $1\, ^1\!F_3$ & 3.053 &  \\
                  & $2\, ^3\!F_3$ & 3.409 &  \\
                  & $2\, ^1\!F_3$ & 3.411 &  \\
                  & $3\, ^3\!F_3$ & 3.733 &  \\
                  & $3\, ^1\!F_3$ & 3.735 &  \\
                  & $4\, ^3\!F_3$ & 4.031 &  \\
                  & $4\, ^1\!F_3$ & 4.035 &  \\
                   \hline
                   \hline
                   & Possible & & \\
                   $J^P=?^?$& assignment(s)&  This work &  Exp.\\
                    \hline
                    $D_{sJ} (3040)^\pm$ &
                   & &3.044\\
                   & $2\, ^3\!P_1\, (1^+)$ & 2.931 & \\
                   & $2\, ^1\!P_1\, (1^+)$ & 2.936 &  \\

\end{tabular} 
\end{ruledtabular}
\end{minipage}
\end{table*}

\begin{figure*}[htb]
\centering
\centering
\includegraphics[width=\linewidth]{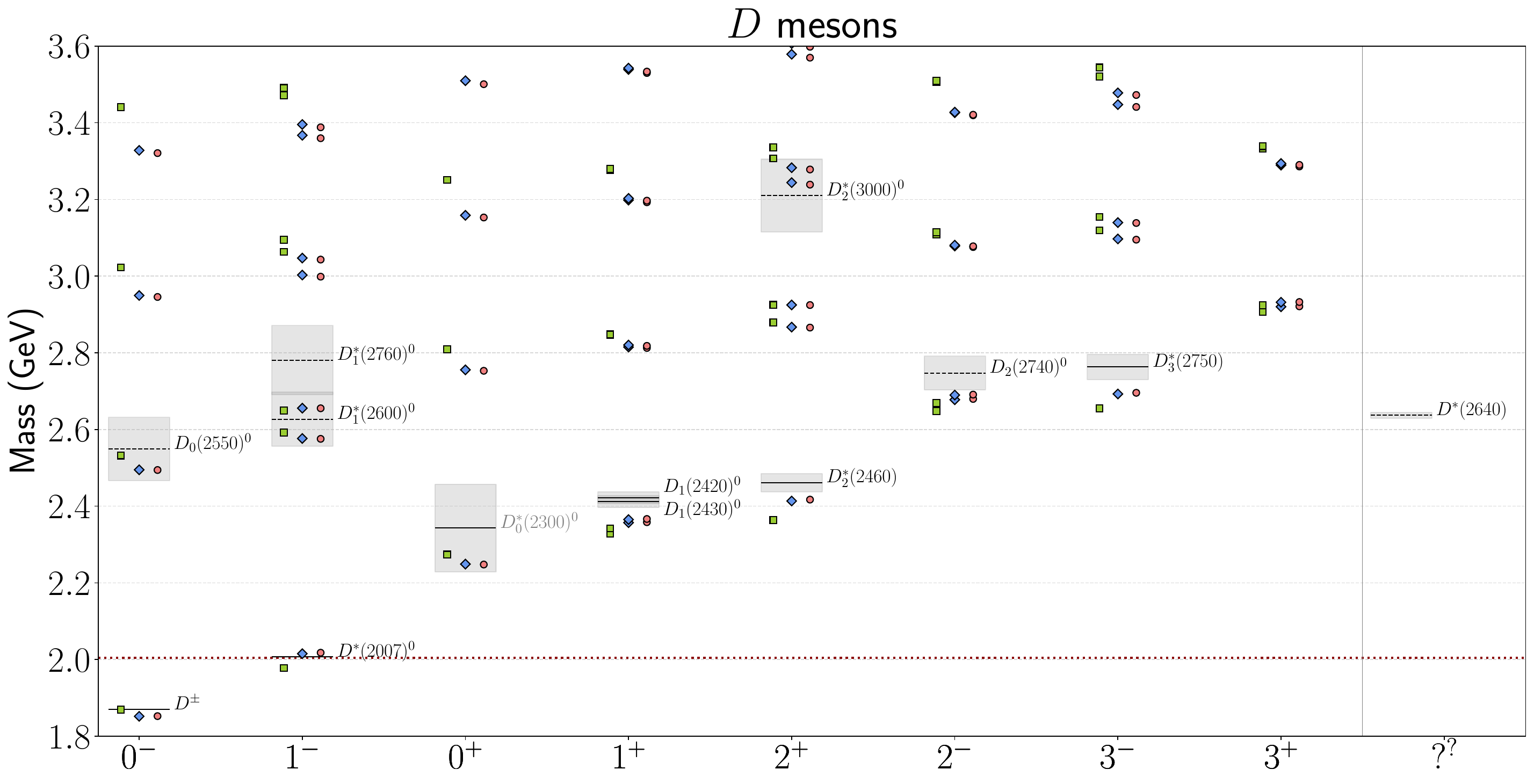}
	\caption{(Color online) Masses of $D$ mesons with $J^P$ from $0^\pm$ to $3^\pm$. Measured states with unknown quantum numbers are shown in an extra column labeled ``$?^?$''. The results of fits to pseudoscalar (green squares), non-axial (blue diamonds),  and to states of all $J^P$ (red circles), are compared with established (solid lines) and unconfirmed (dashed lines) experimental masses, with the grey shading displaying the width. The red dotted line indicates the lowest open-flavor threshold. State names in grey font indicate possible non-$q\bar q$ (exotic) candidates.}
\label{fig:cq}
\end{figure*}

\begin{table*}[t]
\caption{$D$ meson masses. All states are coupled, and the dominant state with quantum numbers $n\, ^{2S+1}\!L_J$ is the one with the highest probability. }
\label{tab:cq}
\begin{minipage}[t]{0.45\textwidth}
\vspace{0pt}
\begin{ruledtabular}
 \begin{tabular}{r|ccc}
  & Dominant & & \\
Meson  & state & This work & PDG  \\
\hline
$J^P=0^-$\\
\hline
$D^\pm$            & $1\, ^1\!S_0$  & 1.852 & 1.8695 \\
$D_0(2550)^0$ & $2\, ^1\!S_0$  & 2.494 & 2.549 \\
               & $3\, ^1\!S_0$  & 2.946 &  \\
               & $4\, ^1\!S_0$  & 3.321 &  \\
               & $5\, ^1\!S_0$  & 3.653 &  \\
               & $6\, ^1\!S_0$  & 3.958 &  \\
  \hline
 $J^P=1^-$\\
\hline
$D^*(2007)^0$  & $1\, ^3\!S_1$ & 2.018 & 2.00685  \\
$D_1^*(2600)^0$ & $2\, ^3\!S_1$ & 2.576 & 2.627 \\
$D_1^*(2760)^0$ & $1\, ^3\!D_1$ & 2.655 & 2.781    \\
                & $3\, ^3\!S_1$ & 2.999 &          \\
                & $2\, ^3\!D_1$ & 3.043 &          \\
                & $4\, ^3\!S_1$ & 3.360 &          \\
                & $3\, ^3\!D_1$ & 3.388 &          \\
                & $5\, ^3\!S_1$ & 3.684 &          \\
                & $4\, ^3\!D_1$ & 3.704 &          \\
                & $6\, ^3\!S_1$ & 3.983 &          \\
 \hline
$J^P=0^+$\\
 \hline
               & $1\, ^3\!P_0$ & 2.248 & \\
$D_0^*(2300)^0$ & & & 2.343 \\
               & $2\, ^3\!P_0$ & 2.753 & \\
               & $3\, ^3\!P_0$ & 3.153 &  \\
               & $4\, ^3\!P_0$ & 3.501 &  \\
               & $5\, ^3\!P_0$ & 3.816 &  \\
\hline
$J^P=1^+$\\
 \hline
 $D_1(2430)^0$ & $1\, ^3\!P_1$ & 2.358 & 2.412 \\
 $D_1(2420)^0$ & $1\, ^1\!P_1$ & 2.367 & 2.4221 \\
             & $2\, ^3\!P_1$ & 2.813 & \\
             & $2\, ^1\!P_1$ & 2.818 &  \\
             & $3\, ^3\!P_1$ & 3.193 &  \\
             & $3\, ^1\!P_1$ & 3.197 &  \\
             & $4\, ^3\!P_1$ & 3.530 &  \\
             & $4\, ^1\!P_1$ & 3.533 &  \\
\end{tabular} 
\end{ruledtabular}
\end{minipage}
\hfill
\begin{minipage}[t]{0.45\textwidth}
\vspace{0pt}
\begin{ruledtabular}
 \begin{tabular}{r|ccc}
  & Dominant & & \\
Meson  & state & This work & PDG  \\
\hline
 $J^P=2^+$\\
 \hline
$D_2^*(2460)$  & $1\, ^3\!P_2$ & 2.417 & 2.4611 \\
                 & $2\, ^3\!P_2$ & 2.866 & \\
                 & $1\, ^3\!F_2$ & 2.925 &  \\
$D^*_2(3000)^0$ & $3\, ^3\!P_2$ & 3.239 & 3.214 \\
                 & $2\, ^3\!F_2$ & 3.278 &  \\
                 & $4\, ^3\!P_2$ & 3.570 &  \\ 
                 & $3\, ^3\!F_2$ & 3.598 &  \\ 
\hline
$J^P=2^-$\\
 \hline
                   & $1\, ^1\!D_2$ & 2.680 & \\
                  & $1\, ^3\!D_2$ & 2.691 &  \\
$D_2(2740)^0$ & & & 2.747 \\
                  & $2\, ^1\!D_2$ & 3.076 &  \\
                  & $2\, ^3\!D_2$ & 3.078 &  \\
                  & $3\, ^1\!D_2$ & 3.419 &  \\
                  & $3\, ^3\!D_2$ & 3.421 &  \\
                  & $4\, ^1\!D_2$ & 3.731 &  \\
                  & $4\, ^3\!D_2$ & 3.734 &  \\
 \hline
$J^P=3^-$\\
 \hline
$D_3^*(2750)$ & $1\, ^3\!D_3$ & 2.696 & 2.7631 \\
                  & $2\, ^3\!D_3$ & 3.095 &  \\
                  & $1\, ^3\!G_3$ & 3.139 &  \\
                  & $3\, ^3\!D_3$ & 3.441 &  \\
                  & $2\, ^3\!G_3$ & 3.472 &  \\
                  & $4\, ^3\!D_3$ & 3.755 &  \\ 
                  & $3\, ^3\!G_3$ & 3.778 & \\
\hline
$J^P=3^+$\\
 \hline
                  & $1\, ^1\!F_3$ & 2.921 &  \\
                  & $1\, ^3\!F_3$ & 2.933 &  \\
                  & $2\, ^1\!F_3$ & 3.286 &  \\
                  & $2\, ^3\!F_3$ & 3.290 &  \\
                  & $3\, ^1\!F_3$ & 3.611 &  \\
                  & $3\, ^3\!F_3$ & 3.613 &  \\
                  & $4\, ^1\!F_3$ & 3.909 &  \\
                  & $4\, ^3\!F_3$ & 3.912 &  \\
                  \hline
                  \hline
                  & Possible & & \\
                  $J^P=?^?$& assignment(s)&  This work &  PDG\\
                  \hline
                  $D^\ast (2640)^\pm$ &
                  & &2.637\\
                  & $2\, ^3\!S_1\,(1^-)$ & 2.576 & \\
                 & $1\, ^3\!D_1\,(1^-)$ & 2.655 &     \\
                 
\end{tabular} 
\end{ruledtabular}
\end{minipage}
\end{table*}

We have solved the GE (\ref{eq:1CSE}) numerically, using expansions of the momentum space wave functions in a basis of B-spline functions \cite{Uzz99} that have been modified for the correct asymptotic behavior, for each meson sector from $b\bar b$ to $c\bar q$ (where $q=u, d$) for the $J^P=0^\pm, 1^\pm, 2^\pm,$ and $3^\pm$ channels. The initially nine model parameters---the three interaction strengths $\sigma$, $\alpha(0)$, $C$, the two cut-off parameters $\lambda_{\rm L}$ and $\lambda_{\rm G}$, and the four constituent quark masses $m_q, m_s, m_c$, and $m_b$ of the light (up and down), strange, charm, and bottom quarks, respectively---were adjusted through global fits to three different sets of experimentally measured states: one set with 10 exclusively pseudoscalar states, another set containing 33 non-axial states ($J^P=0^\pm, 1^-, 2^+, 3^-$), and finally a set with 49 states of all channels (except of $3^+$ where no meson has yet been observed). Our main result is the model fitted to the largest data set, whereas the purpose of the other two models is to see how much of the meson spectrum is already determined by fitting only to smaller subsets of measured data.

Our fits are weighted least-square fits, where lower-mass and well established states are given higher weight than those of higher-mass states. This is mainly due to the expectation that our calculations for higher excitations, especially above the open-flavor thresholds, are less realistic, because no coupling to other possible decay channels has been included.

The parameter values for the resulting three models are listed in Table~\ref{tab1}. Note that the strength $C$ of the constant interaction is omitted from the table because the fits consistently yield $C\approx 0$. In models with a fixed $\alpha_{\mathrm s}$, the constant potential incorporates to some extent the effect of the missing running behavior. By explicitly including the momentum dependence of $\alpha_{\mathrm s}$, the constant interaction becomes redundant, leaving us with only eight adjustable model parameters. It turned out that implementing the running of $\alpha_{\mathrm s}$ substantially improves the global fit compared to using a constant $\alpha_{\mathrm s}$. The best overall agreement is obtained with $N_f=2$ active flavors, which performs slightly better than $N_f=3$, while $N_f=4$ and $N_f=5$ yield significantly poorer fits. Table~\ref{tab1} also shows the extent to which the parameter values depend on the specific set of states to which they are fitted. 

The resulting mass spectra for the three models are presented in Figs.~\ref{fig:bb}-\ref{fig:cq} and Tables \ref{tab:bb}-\ref{tab:cq}. As the form of the wave functions of Eqs.~(\ref{eq:PsiprhoCGSH}) and (\ref{eq:PsipmrhoCGSH}) shows, all our solutions are mixtures of different $\rho$'s and either spin or orbital angular momenta, except for pseudoscalar and scalar mesons where only different $\rho$ states are coupled. Tables \ref{tab:bb}-\ref{tab:cq} show also the ``dominant partial wave'', i.e., the state with the highest probability in the mixture. In a few axial $B_s$, $B$, and charmonium states, the probabilities of spin singlet and triplet states differ by only a few percent. In these cases, both coupled states are given, the first in order having the slightly higher probability.

As discussed in \cite{PhysRevD.96.074007}, in a relativistic framework like CST, both spin singlet an triplet states can contribute to a given $C$-parity of a quarkonium state. 
Because the one-channel CST equation (\ref{eq:1CSE}) is not symmetric under charge conjugation, its solutions for axial quarkonia are only approximate $C$-parity eigenstates. To generate exact $C$-parity eigenstates, a more complicated charge-conjugation--symmetric CST formulation would be required (see Refs.~\cite{mass_function_paper,PhysRevC.26.2203} for details).

We find that fitting only the pseudoscalar states already yields, in most cases, quite accurate predictions for the remaining $J^P$ channels, including the higher-spin tensor states. The most significant deviations from the other two models occur primarily in the higher excited states. Naturally, fitting to the largest set of data provides the best overall description of the spectrum. However, fitting only to non-axial states yields results that are almost as good, as reflected in the similar parameter values listed in Table~\ref{tab1}. 

Across all seven sectors considered in this work, from bottomonium to charmed non-strange mesons, the measured states lying below the respective lowest open-flavor threshold are very well reproduced by our models, as can be seen in Figs.~\ref{fig:bb}-\ref{fig:cq}.
Above threshold, the results of our calculations naturally become less reliable, since they do not account for coupling to other open channels that would introduce mass shifts and decay widths. 

In the following, we discuss our results in detail sector by sector.

\subsubsection{Bottomonium}
In the $b\bar b$ sector, the fit to the pseudoscalar states alone already provides an accurate description of all measured states below threshold, including the tensor states, and yields predictions for several yet-to-be-observed states, as shown in Fig.~\ref{fig:bb}. Table~\ref{tab:bb} compares the masses obtained from our best model---the fit to all states---with the PDG values.

\subsubsection{Bottom-charm mesons}  For the two confirmed $B_c$ states, $B_c^+$ and $B_c(2S)^+$, the $J^P$ quantum numbers listed in the PDG are quark model predictions which are reproduced by our model. We additionally predict several states lying below the lowest open-flavor threshold, as shown in Fig.~\ref{fig:bc}. LHCb has recently reported the first observation of orbitally excited $B_c^+$ states, in the form of two narrow peaks at $6704.8$~MeV and $6752.4$~MeV~\cite{LHCb:2025uce,LHCb:2025ubr}, now included in the PDG as $B_c(1P)^\pm$ states. Their $J^P$ quantum numbers remain undetermined, and we therefore place them in the $J^P=?^?$ column of Fig.~\ref{fig:bc}. As discussed in Ref.~\cite{LHCb:2025uce}, the observed states are consistent with the $\,^3\!P_0$, $\, ^3\!P_1 $, $\, ^1\!P_1$, and $\,^3\!P_2$ assignments. Our corresponding predicted masses---$6700$, $6748$, $6759$, and $6779$~MeV (Table~\ref{tab:bc})---agree with this interpretation to within $5$~MeV, and favor $J^P=0^+$ or $1^+$ for the lower-lying peak at $6704.8$~MeV and $1^+$ or $2^+$ for the higher-lying peak at $6752.4$~MeV. Even more recently, ATLAS announced the first observation of a new state at $6339.0$~MeV decaying into $B_c^+\gamma$, with a significance above $8\sigma$~\cite{ATLAS:2026ubk}. This state is consistent with theoretical expectations for the $B_c^{*+}$ meson---an identification also supported by our prediction of the lowest $1^-$ state at $6361$~MeV. 

\subsubsection{Bottom-strange mesons}
In the $B_s$ sector, the $J^P$ quantum numbers of the four PDG-listed states are quark-model predictions requiring experimental confirmation. The $J^P$ assignments of the two states below threshold are reproduced by our model, and for the $B_{s1}(5830)^0$ ($1^+$) and $B^*_{s2}(5840)^0$ ($2^+$) states above threshold, our predicted masses lie below but close to the PDG values, as shown in Fig.~\ref{fig:bs} and Table~\ref{tab:bs}, confirming the other quark-model assignments. The $B_{sJ}(6063)^0$ and $B_{sJ}(6114)^0$ states, with undetermined $J^P$ quantum numbers, are most likely $L=2$ states~\cite{LHCb:2020pet}, consistent with our predicted masses for the $1\,^3\!D_1$ ($1^-$), $1\,^3\!D_3$ ($3^-$), and the two lowest $2^-$ states, $1\,^3\!D_2$ and $1\,^1\!D_2$. Reference~\cite{LHCb:2020pet} additionally reports alternative  experimental masses for these states, obtained by assuming a decay through $B^{*\pm} K^\mp$. These values are shifted upward by approximately $45$~MeV, leading to $B_{sJ}(6108)^0$ and $B_{sJ}(6158)^0$. Of the two, only the lower one---despite its small experimental uncertainty---remains compatible with an $L=2$ state in our model. Finally, the as-yet-unconfirmed $B^*_{sJ}(5850)$ could correspond to the $1\,^1\!P_1$ ($1^+$) in our spectrum.

\subsubsection{Bottom mesons}
The $B$-meson sector is similar to the $B_s$ case: the $J^P$ assignments of the PDG-listed states are only quark-model predictions, the $B^\pm$ and $B^\ast$ states lying below threshold are well described by our model, and our predictions for the lowest $1^+$ and $2^+$ states lie a bit above the PDG values of $B_{1}(5721)$ and $B^*_{2}(5747)$. Some deviation is expected, since these states lie already significantly above threshold, as shown in Fig.~\ref{fig:bq} and Table~\ref{tab:bq}. Of the unassigned states, $B_J^*(5732)$ has been interpreted as a superposition of several narrow and broad resonances ($B_0^*$, $B_1^*$, $B_1$, and $B_2^*$) in Ref.~\cite{CDF:1999zui} and in quark-model analyses~\cite{PhysRevLett.71.4116, PhysRevD.57.4041,PhysRevD.57.5663}, an interpretation that our model supports. The properties of $B_J(5840)$ and $B_J(5970)$ are consistent with those expected for $B(2S)$ and $B^*(2S)$~\cite{LHCb:2015aaf}. Our $2S$ predictions agree remarkably well with the $B_J(5840)$, while the $B_J(5970)$ is consistent with our $1D$ prediction in the $1^-$ channel. In either case, these states lie too far above the lowest open-flavor threshold for a reliable assignment within our model. 

\subsubsection{Charmonium}
In the charmonium sector, the listed states below and slightly above threshold are well described, including the lowest $2^+$, $2^-$ and $3^-$ tensor states (see Fig.~\ref{fig:cc} and Table~\ref{tab:cc}). Discrepancies appear higher above threshold in the vector, scalar, axial-vector and $2^+$ channels. For several $J=1$ states. such as the vector states $\psi(4230)$, $\psi(4360)$, and $\psi (4660)$, and the axial-vector states $\chi_{c1} (3872)$, $\chi_{c1} (4140)$, and $\chi_{c1} (4685)$, this is not surprising, since they are widely regarded as candidates for exotic structures.

In the scalar channel, the $\chi_{c0}(3860)$ is favored over a $2^{++}$ assignment by $2.5\sigma$ according to Ref.~\cite{Belle:2017egg}, whereas LHCb does not observe this state at all~\cite{LHCb:2020pxc}. Near this mass, our model predicts only a single scalar state, a $\chi_{c0}(2P)$, at $3894$~MeV close to the $\chi_{c0}(3915)$, and a single $2^+$ state, a $\chi_{c2}(2P)$ at $3979$~MeV close to the $\chi_{c2}(3930)$. The latter coincides with the GI-model prediction~\cite{Godfrey1985,PhysRevD.72.054026}. Another quark model using a screened confining potential~\cite{PhysRevD.79.094004} places the $\chi_{c0}(2P)$ at $3842$~MeV and the $\chi_{c2}(2P)$ at $3937$~MeV, yielding a mass splitting in agreement with ours.

The four listed states with unassigned quantum numbers all lie above  threshold. For instance, Ref.~\cite{PhysRevLett.133.131902} reports the measurement of an $\eta_c(3945)$, with $J^{PC}=0^{-+}$, and comments that its mass and width are consistent with those of the  $X(3940)$ at $3942$~MeV. It also speculates that this $\eta_c(3945)$ could be the $\eta_c(3S)$, predicted at $4064$~MeV, of Ref.~\cite{PhysRevD.72.054026}. Our own $\eta_c(3S)$ mass of $4085$~MeV lies even a little higher, so its identification with the new $\eta_c(3945)$ appears somewhat less plausible. An alternative interpretation by Ref.~\cite{Ortega:2017qmg} places the $X(3940)$ in the $1^{++}$ sector as a mixture of charmonium and meson-meson channels. In this scenario, our $2\,^3\!P_1$ prediction at $3934$~MeV would lie within $10$~MeV of the observed mass. However, we do not regard either assignment as conclusive: the discrepancy in the $0^{-+}$ channel may reflect coupled-channel effects absent from our calculations, while the agreement in the $1^{++}$ channel could be coincidental if the assignment is incorrect.

The $X(4160)$ at $4153$~MeV was observed by LHCb with a significance of $4.8\sigma$, with a $2^{-+}$ assignment favored over alternatives by $4\sigma$~\cite{LHCb:2021uow}. Our $2D_2$ states at $4197$ and $4200$~MeV lie a little higher but remain within the large width of the $X(4160)$. 

For the unassigned $X(4350)$, discovered by Belle with a significance of $3.2\sigma$~\cite{Belle:2009rkh}, our model cannot definitively discriminate between $0^+$ and $2^+$. Finally, the $X(4630)$ was reported by LHCb with a significance of $5.5\sigma$. A $J^P = 1^-$ assignment is favored over $2^-$ by $3\sigma$, while all other assignments are disfavored by more than $5\sigma$~\cite{LHCb:2021uow}. Our calculated spectrum would be compatible with either assignment. In any case, the $X(4630)$ is well known to display properties atypical of a conventional $q\bar q$ meson and is therefore a strong candidate for an exotic structure, outside the scope of our model.

\subsubsection{Charm-strange mesons} In the $D_s$ sector, the above-threshold pseudoscalar, vector and tensor ($2^+$ and $3^-$) states are reasonably well described. By contrast, the sub-threshold $D_{s0}^*(2317)^\pm$ is poorly reproduced, likely owing to a significant non-$q\bar{q}$ component in its wave functions. The $D_{s1}(2460)^\pm$ is similarly suspected of carrying a sizeable non-$q\bar q$ admixture. Here our result agrees better with experiment, although this may be a coincidence, but our $D_{s1}(2536)^\pm$ is clearly too low. 

Regarding a possible interpretation of the $D_{sJ}(3040)^\pm$ state at $3044$~MeV, Ref.~\cite{Matsuki_2007} predicts two $1^+$ radial excitations at $3082$~MeV and $3094$~MeV. Our model yields the corresponding $2\,^3\!P_1$ and  $2\,^1\!P_1$ states at somewhat lower values of $2931$~MeV and $2936$~MeV, respectively, though these values lie within the large $240$~MeV width of the $D_{sJ}(3040)^\pm$. 

A more stringent test is offered by the very recent LHCb observation of a new excited charm-strange meson with a statistical significance exceeding $10\sigma$~\cite{LHCb:2026jk}. This state, denoted $D_{s1}(2933)^+$, has a measured mass of $2933$~MeV, a width of $72$~MeV, and a spin-parity $1^+$, making it a natural candidate for a $D_{s1}(2P^{(')})^+$ state. It agrees remarkably well with our two $2P_1$ predictions at $2931$~MeV and $2936$~MeV.

\subsubsection{Charmed mesons} Finally, in the non-strange $D$-meson sector, only the lowest pseudoscalar state lies below the threshold, while the lowest vector state sits right at it. Both masses are well reproduced, whereas our predictions for the above-threshold states lie systematically below the data. The $D_{0}^*(2300)$ ($0^+$) shows a particularly large deviation, likely reflecting its supposed non-$q\bar{q}$ character. 

The two tensor states, $D_{2}(2740)^0$ ($2^-$) and $D_{3}^*(2740)$ ($3^-$), are also of interest. Although they lie well above threshold and our predicted masses are noticeably lower than the measured values, the $2^-$ and $3^-$ assignments remain consistent with our spectrum. 

The $D^*(2640)^\pm$ was reported by DELPHI~\cite{DELPHI:1998oyl} with quantum numbers compatible with $J^P = 1^-$, and its mass would match either our $2\,^3\!S_1$ or $1\,^3\!D_1$ candidate. However, the state has not been confirmed by OPAL~\cite{OPAL:2001xfs} or ZEUS~\cite{ZEUS:2008nzg}, so its existence remains in doubt. 

Finally, the $D^*_2(3000)^0$ with a measured mass of $3214$~MeV, is still omitted from the PDG summary table but appears in the latest PDG listings with $J^P = 2^+$. Our model is consistent with this assignment, with the $3\,^3\!P_2$ state predicted at $3239$~MeV.

\section{Summary and conclusions}\label{sec6}
   In this work, we have generalized the one-channel CST formalism, previously developed for mesons with $J^P=0^\pm$ and $1^\pm$, to states of arbitrary $J^P$. The fully relativistic meson wave functions are expanded in a basis of quark-antiquark states with definite spin $S$ and orbital angular momentum $L$, which are coupled to total angular momentum $J$ using spherical-tensor operators. The use of a spherical-tensor basis constructed from $2\times2$ spin matrices and orbital partial waves facilitates a direct comparison with wave functions obtained in nonrelativistic approaches, in contrast to Bethe-Salpeter formulations that typically expand vertex functions in terms of covariant tensors multiplied by invariant scalar functions~\cite{Krassnigg2011,PhysRevD.84.014014}. The latter are more useful in other applications, such as the calculation of meson form factors. In this work, we have derived, for arbitrary $J^P$, the relations between the Lorentz invariant functions and the spin- and orbital partial waves. We have also improved the previous CST meson model by incorporating the momentum dependence of the strong coupling in the OGE interaction kernel.
   
Our complete kernel has eight adjustable parameters: the strength $\sigma$ of the linear confining interaction, a regulator $\tau$ that fixes the value of $\alpha_{\mathrm{s}}(q^2)$ at $q^2=0$ in the OGE interaction, the constituent masses of the light, strange, charm, and bottom quarks, and two Pauli-Villars cut-off parameters. Using this model, we have calculated the masses and wave functions of heavy and heavy-light mesons, from the $D$-meson sector to bottomonium, for all $J^P$ channels with established states listed by the Particle Data Group, i.e., from $J=0$ up to $J=3$ for both natural and unnatural parities.    
   
The model parameters were adjusted through global fits of the masses to the experimental data. 
We have constructed three models that differ only in the set of experimental masses to which they were fitted. It is remarkable that a fit performed only to pseudoscalar states already yields accurate predictions for the remaining $J^P$ channels, including the higher-spin tensor states. Naturally, the fit to the full set of available states provides the best overall description of the spectrum. However, it is quite similar to that obtained from fitting to all except the unnatural parity states $1^+, 2^-, $ and $3^+$, and it does not differ dramatically from that of the pseudoscalar fit. 
Differences between the masses calculated from the three models become more noticeable primarily for higher excited states. 

These results show that our previous findings~\cite{Leit_o_2017,PhysRevD.96.074007} apply also to tensor mesons: the covariant structure of the kernel correctly captures the essential part of the spin dependence of the quark-antiquark interaction, which also allows accurate predictions of higher-spin states. Furthermore, we find that the inclusion of a running strong coupling removes the need for a constant interaction in the kernel and improves the predicted meson spectra. The most accurate unified description of the mass spectrum for $q \bar q$ mesons containing at least one heavy quark is achieved when two active flavors ($N_f=2$) are assumed.
   
   We have found several states in the spectrum that are not well described by our model. Most of them are known---or at least suspected---to contain exotic non-$q\bar{q}$ structures, or are high above threshold. In addition, we have proposed $J^P$ assignments for several states, some of which only recently observed, whose quantum numbers have not yet been experimentally determined. In the bottom-charm sector, our predictions are consistent with the LHCb $B_c(1P)^+$ candidates at $6704.8$ and $6752.4$~MeV\cite{LHCb:2025uce}, and identify the ATLAS state at $6339$~MeV~\cite{ATLAS:2026ubk} as the  $B_c^{*+}$. In the charm-strange sector, the recently observed $D_{s1}(2933)^+$~\cite{LHCb:2026jk} is in close agreement with our two $2\,P_1$ predictions, supporting its interpretation as a $D_{s1}(2P^{(')})^+$ state. 
 
   Despite its overall success, the present formulation has still some limitations: For instance, exact charge-conjugation invariance has not yet been implemented, and the one-channel CST solutions for quarkonia with unnatural parity and $J \ge 1$ are therefore only approximate $C$-parity eigenstates. Incorporating this symmetry requires at least a two-channel CST formulation \cite{mass_function_paper}, which is more difficult to implement and will be solved at a later time.
   
Another refinement of our meson spectrum calculations that we intend to implement in the near future is the inclusion of  a momentum-dependent mass function in the light-quark propagator. Throughout the present work, we have assumed constant constituent quark masses in the quark propagators. While this approximation is reasonable for heavy quarks, where the effect of dynamical mass generation is small, the running of the quark mass is expected to be important for light and strange quarks. Therefore we expect the replacement of constant quark masses by mass functions to have a noticeable impact on the spectra of heavy-light mesons.

   \begin{acknowledgments}
   	This work was supported by FCT under grant numbers CERN/FIS-PAR/0023/2021 and  UID/04349/2025 (https://doi.org/10.54499/UID/04349/2025). 
   \end{acknowledgments}

   \appendix

 \section{Construction of the $K^\rho_j$ operators}
 \label{eq:derivationKrhoj}
 Here we derive the explicit expressions of the $K^\rho_j$ for natural and unnatural parity states.
 The CST wave functions are expanded in irreducible spherical tensors of total angular momentum $J$, constructed by coupling orbital angular momentum $L$ and total spin 
 $S$ with the four possible cases (i)-(iv)  discussed in Sec.\ \ref{sec3}.
 The irreducible tensors are obtained from the tensor product of orbital and spin tensors,
 \begin{eqnarray}
 	T^{(L,S)}_{m_J}(\hat{\boldsymbol p})&=& \left\{T^{(L)}_{m_L}(\hat{\boldsymbol p})\otimes T^{(S)}_{m_S}\right\}_{m_J}^{(J)}\nonumber\\&=&\sum_{m_L=-L}^L\sum_{m_S=-S}^S T^{(L)}_{m_L} (\hat{\boldsymbol p})  T^{(S)}_{m_S}  C_{Lm_LS m_S}^{Jm_J}\, \nonumber\\   \label{eq:TSL}
 \end{eqnarray}
 where $C$ are the Clebsch-Gordan coefficients. Our goal is to express the coupled tensors $T_{m_J}^{(L,S)}$ explicitly in terms of the basic building blocks: the three-momentum direction $\hat {\boldsymbol p}$, the Pauli matrices $\boldsymbol {\sigma}$ and the spin-1 polarization vectors $\boldsymbol {\xi}$.
 
 The orbital tensor is built from the tensor product of $L$ tensors $T_\lambda^{(1)}$ with orbital angular momentum 1:
 \begin{widetext}  
 	\begin{eqnarray}
 		T^{(L)}_{m_L}(\hat{\boldsymbol p})&\equiv&
 		Y_{L m_L}(\hat{\boldsymbol p}) \nonumber\\&=& \sqrt{(4\pi)^{L-1}}\sqrt{\frac{(2L+1)!!}{L!\,3^L}} \left\lbrace T^{(1)}_{\lambda_1}(\hat{\boldsymbol p}) \otimes T^{(1)}_{\lambda_2}(\hat{\boldsymbol p})\otimes\cdots\otimes T^{(1)}_{\lambda_L}(\hat{\boldsymbol p}) \right\rbrace_{m_L}^{(L)}
 		\nonumber\\&=&
 		\frac{1}{\sqrt{4\pi}}\sqrt{\frac{(2L+1)!!}{L!}}
 		\sum_{ \lambda_1=-1}^1 \sum_{ \lambda_2=-1}^1 \cdots  \sum_{ \lambda_L=-1}^1\sum_{ \lambda_{12}=-2}^2 \sum_{ \lambda_{123}=-3}^3 \cdots \sum_{\lambda_{12\ldots L-1}=-(L-1)}^{L-1} \hat p_{\lambda_1}\hat p_{\lambda_2}\cdots \hat p_{\lambda_L} \nonumber\\&&\times C^{2\lambda_{12}}_{1\lambda_11\lambda_2}C^{3\lambda_{{123}}}_{2\lambda_{12}1\lambda_3}\cdots C^{Lm_L}_{(L-1)\lambda_{12\ldots L-1}1\lambda_L}\,\label{eq:orbitaltensor}
 	\end{eqnarray}
 \end{widetext} 
 where
 \begin{eqnarray}
 	T^{(1)}_{\lambda}(\hat{\boldsymbol p})\equiv Y_{1 \lambda}(\hat{\boldsymbol p})=\sqrt{\frac{3}{4\pi}}\hat p_{\lambda} \,\quad \text{with}\quad \lambda=-1,0,1 \, , \nonumber\\
 \end{eqnarray}
 and $\hat p_{\lambda}$ is a spherical component of $\hat{\boldsymbol p}=\boldsymbol p/p$, defined by\footnote{The spherical component $\hat p_{\lambda}$ of the relative unit three-vector $\hat{\boldsymbol p}$ should not be confused with the on-shell four-vector $\hat p_1$ of quark~1.} 
 \begin{eqnarray}
 	\hat p_{\pm 1}=\mp \frac1{\sqrt 2 }\frac{p_x\pm \mathrm i\,p_y}{p}\,,\quad \hat p_{0}=\frac{p_z}{p}\,.
 \end{eqnarray}
 The spherical tensors are orthonormal under angular integration which fixes the normalization factor in front of the tensor products in~(\ref{eq:orbitaltensor}):  \begin{eqnarray}
 	\int d \Omega(\hat {\boldsymbol{p}}) \left[T^{(L')}_{m_L'}(\hat {\boldsymbol{p}})\right]^\ast T^{(L)}_{m_L}(\hat {\boldsymbol{p}})=\delta_{LL'}\delta_{m_L'm_L}\,.
 \end{eqnarray}
 The spin tensor is obtained as the tensor product of two spin-$\frac12$ spherical tensors,
 \begin{eqnarray}
 	T_{m_S}^{(S)}&=&\left\lbrace T^{(\frac12)}_{\lambda_1} \otimes \left( T^{(\frac12)}_{\lambda_2}\right)^\dag \right\rbrace_{m_S}^{(S)}
 	\nonumber\\&=&
 	\sum_{\lambda_1=-\frac12}^{\frac12}\sum_{\lambda_2=-\frac12}^{\frac12}\chi^{}_{\lambda_1} \chi_{\lambda_2}^\dag C_{\frac12\lambda_1\frac12\lambda_2}^{Sm_S}\nonumber\\&=&
 	\begin{cases}
 		\frac{\mathrm i}{\sqrt{2}}\sigma^2\,, & \text{if } S=0\,; \\
 		\frac{\mathrm i}{\sqrt{2}}\sigma_{m_S}\sigma^2\,, & \text{if }S=1\,,
 	\end{cases}\label{eq:spintensor}
 \end{eqnarray}
 with $ T^{(\frac12)}_{\lambda_i}\equiv\chi^{}_{\lambda_i}$. In Eq.~(\ref{eq:spintensor}), $\sigma_{m_S}$ represents the spherical components of the Pauli matrices $\boldsymbol{\sigma}=(\sigma^1,\sigma^2,\sigma^3)$. The spherical spin tensors are themselves $(2\times 2)$ matrices and are orthonormal with respect to the trace, 
 
 \begin{eqnarray}
 	{\rm Tr}\left[ \left(T^{(S')}_{m_S'}\right)^\dag T^{(S)}_{m_S}\right]=\delta_{SS'}\delta_{m_Sm_S'}\,.
 \end{eqnarray}
 
 Before substituting the orbital tensors (\ref{eq:orbitaltensor}) and the spin tensors (\ref{eq:spintensor}) into Eq.~(\ref{eq:TSL}), it is useful to note that 
 \begin{eqnarray}
 	\hat{\boldsymbol{p}}\cdot \boldsymbol \xi_\lambda=\hat{p}^i  \xi^i_\lambda\equiv \hat p_\lambda\,, \quad  \boldsymbol{\sigma}\cdot \boldsymbol {\xi}_\lambda=\sigma^i\xi^i_\lambda\equiv \sigma_\lambda\, ,
 \end{eqnarray}
 where the $\boldsymbol{\xi}_\lambda$ are the spherical spin-1 polarization vectors for massive particles,
 \begin{eqnarray}
 	\boldsymbol{\xi}_{\pm1}=\mp\frac{1}{\sqrt2}(1,\pm\mathrm i,0), \quad  \boldsymbol{\xi}_0=(0,0,1)\,.
 \end{eqnarray}
 For the spin singlet case we obtain
 \begin{widetext} 
 	\begin{eqnarray}
 		T^{(J,0)}_{m_J}(\hat{\boldsymbol p})&=& \frac1{\sqrt 2} \frac{1}{\sqrt{4\pi}}\sqrt{\frac{(2J+1)!!}{J!}}\sum_{m_L=-J}^J  \sum_{ \lambda_1=-1}^1 \sum_{ \lambda_2=-1}^1 \cdots  \sum_{ \lambda_J=-1}^1\sum_{ \lambda_{12}=-2}^2 \sum_{ \lambda_{123}=-3}^3 \cdots \sum_{\lambda_{12\ldots J-1}=-(J-1)}^{J-1} \hat p_{\lambda_1}\hat p_{\lambda_2}\cdots \hat p_{\lambda_J}\mathrm i \,\sigma^2 \nonumber\\&&\times C^{2\lambda_{12}}_{1\lambda_11\lambda_2}C^{3\lambda_{{123}}}_{2\lambda_{12}1\lambda_3}\cdots C^{Jm_L}_{(J-1)\lambda_{12\ldots J-1}1\lambda_J} C_{Jm_L00}^{Jm_J}\nonumber\\
 		&\equiv& \frac1{\sqrt 2} \frac{1}{\sqrt{4\pi}}\sqrt{\frac{(2J+1)!!}{J!}} \hat p^i\hat p^j\cdots \hat p^n\zeta_{m_J}^{ij\ldots n}\, \mathrm i \,\sigma^2\label{eq:T0J}
 		\,  .  \end{eqnarray}
 	In the last step we have introduced the rank-$J$ polarization tensor 
 	\begin{eqnarray}
 		\zeta^{ij\ldots n}_{m_J}&=&
 		\sum_{ \lambda_1 \lambda_2 \ldots  \lambda_{12\ldots J-1}} \xi^{i}_{\lambda_1}\xi^j_{\lambda_2}\cdots \xi^{n}_{\lambda_J} C^{2\lambda_{12}}_{1\lambda_11\lambda_2}\cdots C^{Jm_J}_{(J-1)\lambda_{12\cdots J-1}1\lambda_J}\,\,\label{eq:zetaijk}
 	\end{eqnarray}
 	which collects all Clebsch-Gordan coefficients and polarization vectors. 
 	In an analogous manner, the three tensors corresponding to the spin-triplet configurations can be constructed. Their explicit forms are
 	\begin{eqnarray}
 		T^{(J,1)}_{m_J}(\hat{\boldsymbol p})&=& \frac1{\sqrt 2} \frac{1}{\sqrt{4\pi}}\sqrt{\frac{(2J+1)!!}{J!}}\sum_{m_L=-J}^J \sum_{m_S=-1}^1 \sum_{ \lambda_1=-1}^1 \sum_{ \lambda_2=-1}^1 \cdots  \sum_{ \lambda_J=-1}^1\sum_{ \lambda_{12}=-2}^2 \sum_{ \lambda_{123}=-3}^3 \cdots \! \! \! \! 
		\sum_{\lambda_{12\ldots J-1}=-(J-1)}^{J-1} \! \! \! \! \! \! \! \!  \! \! \! \! \! \! 
		\hat p_{\lambda_1}\hat p_{\lambda_2}\cdots \hat p_{\lambda_J} \sigma_{m_S}\mathrm i \,\sigma^2\nonumber\\&&\times C^{2\lambda_{12}}_{1\lambda_11\lambda_2}C^{3\lambda_{{123}}}_{2\lambda_{12}1\lambda_3}\cdots C^{Jm_L}_{(J-1)\lambda_{12\ldots J-1}1\lambda_J}  C_{Jm_L1m_S}^{Jm_J}\nonumber\\
 		&\equiv&
 		-\frac{1}{\sqrt{2}}\frac{1}{\sqrt{4 \pi}} \sqrt{\frac{J(2J+1)!!}{(J+1)!}} \underbrace{\hat p^{i} \hat p^{j}\cdots \hat p^{m}}_{(J-1) \text{\,factors}} \left(\boldsymbol{\sigma}\cdot \hat {\boldsymbol p}\, \sigma^n - \hat p^{n}\right)\zeta_{m_J}^{ij\ldots mn}\mathrm i \,\sigma^2\label{eq:T1J}
 		\,   ,\end{eqnarray}
  	\begin{eqnarray}
 		T^{(J-1,1)}_{m_J}(\hat{\boldsymbol p})&=& \frac1{\sqrt 2} \frac{1}{\sqrt{4\pi}}\sqrt{\frac{(2J-1)!!}{(J-1)!}}\sum_{m_S=-1}^1\sum_{m_L=-(J-1))}^{J-1}  \sum_{ \lambda_1=-1}^1 \sum_{ \lambda_2=-1}^1 \cdots  \sum_{ \lambda_J=-1}^1\sum_{ \lambda_{12}=-2}^2 \cdots \! \! \! \! 
		\sum_{\lambda_{12\ldots J-2}=-(J-2)}^{J-2}  \! \! \! \! \! \! \! \!  \! \! \! \! \! \! 
		\hat p_{\lambda_1}\hat p_{\lambda_2}\cdots \hat p_{\lambda_{J-1}}  \sigma_{m_S}\mathrm i \,\sigma^2\nonumber\\&&\times C^{2\lambda_{12}}_{1\lambda_11\lambda_2}C^{3\lambda_{{123}}}_{2\lambda_{12}1\lambda_3}\cdots C^{(J-1)m_L}_{(J-2)\lambda_{12\ldots J-2}1\lambda_{J-1}} C_{(J-1)m_L1m_S}^{Jm_J}\nonumber\\
 		&\equiv&
 		\frac1{\sqrt 2} \frac{1}{\sqrt{4\pi}}\sqrt{\frac{(2J-1)!!}{(J-1)!}} \underbrace{\hat p^{i} \hat p^{j}\cdots \hat p^{m}}_{(J-1) \text{\,factors}}  \sigma^n \zeta_{m_J}^{ij\ldots mn} \mathrm i \,\sigma^2\label{eq:T1Jm1}
 		\,   , \end{eqnarray}
 	and
 	\begin{eqnarray}
 		T^{(J+1,1)}_{m_J}(\hat{\boldsymbol p})&=& \frac1{\sqrt 2} \frac{1}{\sqrt{4\pi}}\sqrt{\frac{(2J+3)!!}{(J+1)!}}\sum_{m_S=-1}^1\sum_{m_L=-(J+1)}^{J+1}  \sum_{ \lambda_1=-1}^1 \sum_{ \lambda_2=-1}^1 \cdots  \sum_{ \lambda_J=-1}^1\sum_{ \lambda_{12}=-2}^2  \cdots  \! \!\! \!
		\sum_{\lambda_{12\ldots J}=-J}^{J}  \! \! \! \! \! \! 
		\hat p_{\lambda_1}\hat p_{\lambda_2}\cdots \hat p_{\lambda_{J+1}}  \sigma_{m_S}\mathrm i \,\sigma^2\nonumber\\&&\times C^{2\lambda_{12}}_{1\lambda_11\lambda_2}C^{3\lambda_{{123}}}_{2\lambda_{12}1\lambda_3}\cdots C^{J\lambda_{12\ldots J}}_{(J-1)\lambda_{12\ldots J-1}1\lambda_{J}} C^{(J+1)m_L}_{J\lambda_{12\ldots J}1\lambda_{J+1}} C_{(J+1)m_L1m_S}^{Jm_J}\nonumber\\
 		&\equiv&
 		\frac{1}{\sqrt{2}}\frac{1}{\sqrt{4 \pi}} \sqrt{\frac{(2J-1)!!}{(J+1)!}} \underbrace{\hat p^{i} \hat p^{j} \cdots \hat p^{m}}_{(J-1) \text{\,factors}}\left[-(2J+1)\boldsymbol{\sigma}\cdot \hat {\boldsymbol p}\, \hat p^n +J \sigma^n\right] \zeta_{m_J}^{ij\ldots mn}\mathrm i \,\sigma^2 \label{eq:T1Jp1}
 		\,   . \end{eqnarray}
 \end{widetext}
 These expressions are valid for any $J\geq1$. For the special case $J=0$, the tensors of (\ref{eq:T1J}) and (\ref{eq:T1Jm1}) vanish.
 
 The four tensor structures are orthonormal under angular integration combined with the spin trace, i.e.,
 \begin{eqnarray}
 	\mathrm {Tr}\int \mathrm d \Omega (\hat {\boldsymbol{p}}) \left[T_{m_J'}^{(L',S')}(\hat {\boldsymbol{p}})\right]^\dag T_{m_J}^{(L,S)}(\hat {\boldsymbol{p}})=\delta_{LL'} \delta_{SS'} \delta_{m_Jm_J'} \,.\nonumber\\\label{eq:Tortonorm}
 \end{eqnarray}
 Thus, the $K^\rho_j$'s are proportional to the tensors $T^{(L,S)}_{m_J}$, with normalization fixed by convention. We adopt the Condon-Shortley phase convention for spherical harmonics and Clebsch-Gordan coefficients.\footnote{Compared to Ref.~\cite{PhysRevD.96.074007}, the definition of $K_2^\rho$ differs by an overall sign in front of $T^{(J+1,1)}_{m_J}$.
 	This modification is purely conventional and chosen to allow the CST wave function to be expressed as a clean sum over orbital angular momenta without introducing additional minus signs, see Eqs.~(\ref{eq:PsiprhoCGSH}) and~(\ref{eq:PsipmrhoCGSH}). For readability, we also suppress the explicit dependence of $K_j^\rho$ on the spin polarization $m_j$.} Further note that the roles of the $\rho=+$ and $\rho=-$ components of the $K^\rho_j$'s are interchanged between natural- and unnatural-parity mesons.
 Finally, the orthogonality relation~(\ref{eq:Tortonorm}) (for equal spin projections $m_J=m_J'$) then translates into the following condition for the $K^\rho_j$:
 \begin{eqnarray}
 	\mathrm {Tr}\int \mathrm d \Omega (\hat {\boldsymbol{p}}) \left[K_{j'}^{\rho'}(\hat {\boldsymbol{p}})\right]^\dag K_{j}^{\rho}(\hat {\boldsymbol{p}})=8\pi\delta_{jj'} \delta_{\rho\rho'}  \,.
	\label{eq:Kortonorm}
 \end{eqnarray}
 \section{Covariant-tensor basis}\label{sec4}
 In this section, we introduce a Lorentz-covariant tensor basis for the vertex function $\Gamma(\hat p_1,p_2)$. Other approaches, such as the Dyson-Schwinger/Bethe-Salpeter formalism of Refs.~\cite{LLEWELLYNSMITH1969521,PhysRevD.84.014014,Krassnigg2011}
 typically work directly with a covariant tensor basis where the tensors are transverse to the meson four-momentum $P^\mu$. In the rest frame of the meson, transverse tensors reduce to a three-dimensional form.  
 In our approach, we define an invariant vertex function as the contraction of a transverse polarization tensor with independent covariant tensor structures, each multiplied by some invariant function. For mesons with angular momentum $J$ and natural parity $P=(-1)^J$, the vertex function is a Lorentz scalar written as the contraction of two rank-$J$ Lorentz tensors, 
 \begin{widetext}
 	\begin{eqnarray}
 		\Gamma_{m_J}^{(J)}(\hat p_1,p_2)&=&\zeta^{\mu\nu\sigma\ldots\omega}_{m_J} \Gamma_{\mu\nu\sigma\ldots\omega} (\hat p_1,p_2)\nonumber\\
 		&=&\zeta^{\mu\nu\sigma\ldots\omega}_{m_J} 
 		\left\{G_1 M_{\mu\nu\sigma\ldots\omega}  +G_2N_{\mu\nu\sigma\ldots\omega} +
 		\left[G_3M_{\mu\nu\sigma\ldots\omega}  +G_4N_{\mu\nu\sigma\ldots\omega} \right]\Lambda (-p_2)\right\}\label{eq:tensorGammamunu}\,.
 	\end{eqnarray} 
 	For mesons with unnatural parity $P=(-1)^{J+1}$, the vertex function is a Lorentz pseudoscalar written as the contraction of a tensor and an axial tensor and is given by 
 	\begin{eqnarray}
 		\Gamma^{5(J)}_{m_J}(\hat p_1,p_2) &=&\zeta^{\mu\nu\sigma\ldots\omega}_{m_J} \Gamma^5_{\mu\nu\sigma\ldots\omega}(\hat p_1,p_2) \nonumber\\
 		&=&\zeta^{\mu\nu\sigma\ldots\omega}_{m_J}
 		\left\{G_1 M_{\mu\nu\sigma\ldots\omega} \gamma^5 +G_2N_{\mu\nu\sigma\ldots\omega}\gamma^5 +
 		\left[G_3M_{\mu\nu\sigma\ldots\omega} \gamma^5 +G_4N_{\mu\nu\sigma\ldots\omega}\gamma^5 \right]\Lambda (-p_2)\right\}\label{eq:tensorGamma5munu}\,.
 	\end{eqnarray} 
 \end{widetext}
In these expressions, $\zeta^{\mu\nu\sigma\ldots}_{m_J}$
 is the covariant generalization of the rank-$J$ polarization tensor associated with angular momentum $J$ and spin-polarization $m_J=-J,\ldots,J$. It is obtained from the polarization tensor $\zeta^{ij\ldots n}_{m_J}$ of (\ref{eq:zetaijk}) by replacing the three-vector components $\xi^j_{\lambda_i}$ in $\zeta^{ij\ldots n}_{m_J}$ with the corresponding four-vector components $\xi^{\mu}_{\lambda_i}$. The explicit construction and properties of $	\zeta^{\mu\nu\sigma\ldots\omega}_{m_J}$ are discussed in Section~\ref{app:zeta}.
  The functions $G_i=G_i(\hat p_1\cdot p_2,p_2^2)$ are Lorentz-invariant, while $\Gamma_{\mu\nu\sigma\ldots}(\hat p_1,p_2)$,  $M^{\mu\nu\sigma\ldots}=M^{\mu\nu\sigma\ldots}(p) $ and $N^{\mu\nu\sigma\ldots}=N^{\mu\nu\sigma\ldots}(p)$ are rank-$J$ covariant tensors. Additionally, $\Gamma^5_{\mu\nu\sigma\ldots}(\hat p_1,p_2)$ represents an axial tensor of rank $J$. In Section~\ref{app:Gamma}, we provide explicit expressions for the tensors $M^{\mu\nu\sigma\ldots}$ and $N^{\mu\nu\sigma\ldots}$. 
 \subsection{Rank-$J$ polarization tensor}\label{app:zeta}
 The rank-$J$ spherical polarization tensor $\zeta^{(J)}_{m_J}$ for angular momentum $J$ and projection $m_J=-J,\ldots, J$ can be constructed from the tensor products of $J$ irreducible rank-1 tensors $\xi^{(1)}_{\lambda_i}$, which we abbreviate as $\xi_{\lambda_i}$ in the following: 
 \begin{widetext}
 	\begin{eqnarray}
 		\zeta^{(J)}_{m_J}&=&\left\lbrace \xi_{\lambda_1}\otimes \xi_{\lambda_2}\otimes\cdots\otimes \xi_{\lambda_J} \right\rbrace_{m_J}^{(J)}\nonumber\\&\equiv&
 		\sum_{ \lambda_1=-1}^1 \sum_{ \lambda_2=-1}^1 \cdots  \sum_{ \lambda_J=-1}^1\sum_{ \lambda_{12}=-2}^2 \sum_{ \lambda_{123}=-3}^3 \cdots \sum_{ \lambda_{12\cdots (J-1)}=-(J-1)}^{J-1} \xi_{\lambda_1}\xi_{\lambda_2}\xi_{\lambda_3}\cdots \xi_{\lambda_J} C^{2\lambda_{12}}_{1\lambda_11\lambda_2}C^{3\lambda_{{123}}}_{2\lambda_{12}1\lambda_3}\cdots C^{Jm_J}_{(J-1)\lambda_{12\cdots (J-1)}1\lambda_J}\,.\nonumber\\
 	\end{eqnarray}
 	Here,  the $\xi_{\lambda_i}$'s with $\lambda_i=-1,0,1$ represent the three spherical polarization four-vectors for massive spin-1 particles corresponding to spin polarizations $-1,0,$ and $1$. These vectors, expressed in terms of Lorentz four-vector components, are denoted $\xi_{\lambda_i}^\mu$ where $\mu=0,\ldots,3$. The spherical tensor $\zeta^{(J)}_{m_J}$ can then be written in manifestly Lorentz-covariant form with components\footnote{For easier readability, we omit the superscript $(J)$ from this point onward.}
 	\begin{eqnarray}
 		\zeta^{\mu\nu\sigma\cdots\omega}_{m_J}&=&
 		\sum_{ \lambda_1 \lambda_2 \ldots  \lambda_{12\cdots J-1}} \xi^{\mu}_{\lambda_1}\xi^\nu_{\lambda_2}\xi^\sigma_{\lambda_3}\cdots \xi^{\omega}_{\lambda_J} C^{2\lambda_{12}}_{1\lambda_11\lambda_2}C^{3\lambda_{123}}_{2\lambda_{12}1\lambda_3}\cdots C^{Jm_J}_{(J-1)\lambda_{12\cdots J-1}1\lambda_J}\,\nonumber\\
 		&=&
 		\sum_{ \lambda_1 \lambda_{12} \ldots  \lambda_{12\cdots J-1}} \xi^{\mu}_{\lambda_1}\xi^\nu_{\lambda_{12}-\lambda_1}\xi^\sigma_{\lambda_{123}-\lambda_{12}}\cdots \xi^{\omega}_{\lambda-\lambda_{12\cdots J-1}}  C^{2\lambda_{12}}_{1\lambda_11(\lambda_{12}-\lambda_1)}C^{3\lambda_{123}}_{2\lambda_{12}1(\lambda_{123}-\lambda_{12})}\cdots C^{Jm_J}_{(J-1)\lambda_{12\cdots J-1}1(\lambda-\lambda_{12\cdots J-1})}\, ,\nonumber\\\label{eq:zeta}
 	\end{eqnarray}
 \end{widetext}
 where we have carried out the sums over $\lambda_2,\ldots,\lambda_J$ using the constraints
 \begin{eqnarray}
 	&&\lambda_1+\lambda_2=\lambda_{12}\,,\nonumber\\ &&\lambda_{12}+\lambda_3=\lambda_{123}\,,\nonumber\\\nonumber&&\qquad\vdots\\
 	&&\lambda_{12\cdots J-1}+\lambda_J=m_J\,. \nonumber\end{eqnarray}
 Note that Eq.~(\ref{eq:zeta}) refers to a specific (but arbitrary) choice of coupling order 
 \begin{eqnarray}
 	\boldsymbol J=\underbrace{\Big\{\underbrace{\big[\underbrace{(\underbrace{\boldsymbol j_1+ \boldsymbol j_2}_{\boldsymbol j_{12}})+ \boldsymbol j_3 }_{ \boldsymbol j_{123}}\big]+ \boldsymbol j_4}_{\boldsymbol j_{1234}}\Big\}+\cdots}_{\boldsymbol j_{12\cdots J-1}}+ \boldsymbol j_J\,\nonumber
 \end{eqnarray}
 of the $J$ angular momenta $\boldsymbol j_i$ of magnitude 1. 
  
 In this work, we operate in the meson rest frame, where $P=(\mu,\boldsymbol 0)$, using the standard representation
 \begin{eqnarray}
 	\xi_{\pm1}=\mp\frac{1}{\sqrt2}(0,1,\pm \mathrm i,0), \quad  \xi_0=(0,0,0,1)\,.
 \end{eqnarray}
 Since the $\xi_{\lambda_i}$'s are irreducible spherical tensors of rank 1, $\zeta^{\mu\nu\sigma\cdots}_{m_J}$, the $J$-fold tensor product of $\xi_{\lambda_i}$'s, is an irreducible spherical tensor of rank $J$.  
 Because of the symmetry property of the Clebsch-Gordan coefficients, we have
 \begin{eqnarray}
 	C^{Jm_J}_{j_2\lambda_2 j_1\lambda_1}=(-1)^{j_1+j_2-J}C^{Jm_J}_{j_1\lambda_1j_2\lambda_2}\,,
 \end{eqnarray}
 which implies
 $\zeta^{\mu\nu\rho\cdots}_{m_J}$ is totally symmetric in the Lorentz indices. It also satisfies the transversality condition
 \begin{eqnarray}
 	P_\mu\zeta^{\mu\nu\ldots\omega}_{m_J}=P_\nu\zeta^{\mu\nu\ldots\omega}_{m_J}=\ldots=P_\omega\zeta^{\mu\nu\ldots\omega}_{m_J}=0\,,\label{eq:transverse}\nonumber\\
 \end{eqnarray}
 and the tracelessness condition \begin{eqnarray}
 	\mathrm g_{\mu\nu}\zeta^{\mu\nu\rho\ldots\tau\omega}_{m_J}=\mathrm g_{\mu\rho}\zeta^{\mu\nu\rho\ldots\tau\omega}_{m_J}=\ldots=g_{\tau\omega}\zeta^{\mu\nu\rho\ldots\tau\omega}_{m_J}=0\,.\nonumber\\\label{eq:tracelessness}
 \end{eqnarray}
 The transversality property (\ref{eq:transverse}) follows from the transversality of the massive spin-1 polarization vectors, $P_\mu\xi^\mu_{m_J}=0$, while the tracelessness property (\ref{eq:tracelessness}) follows from the relations 
 \begin{eqnarray}
 	&&\xi_{\mu\pm1} \xi^\mu_{\pm1}=0\,,\quad \label{eq:xipmxipm}\\
 	&&\xi_{\mu\pm1} \xi^\mu_0=\xi_{\mu0} \xi^\mu_{\pm1}=0\,,\label{eq:xipmxi0}\\
 	&&\xi_{\mu\pm1} \xi^\mu_\mp=\xi_{\mu\mp1} \xi^\mu_{\pm1}=1\,,\label{eq:xipmximp}\\
 	&&\xi_{\mu0} \xi^\mu_0=-1\,.\label{eq:xi0xi0}
 \end{eqnarray}
 Tracelessness can be shown as follows: For $m_J=\pm J$, Eq.~(\ref{eq:xipmxipm}) ensures the tracelessness. For $|m_J|=J-1$, the combination involving Eq.~(\ref{eq:xipmxi0}) appear in the trace, ensuring tracelessness in combination with Eq.~(\ref{eq:xipmxipm}). For $|m_J|<J-1$ the combination $\xi_0\cdot \xi_0+ \xi_{\mp1}\cdot \xi_{\pm1}$ appears in the trace, which vanishes because of Eqs.~(\ref{eq:xipmximp}) and~(\ref{eq:xi0xi0}), thereby ensures tracelessness.
 
 For the specific case $J=2$, we have
 \begin{widetext}
 	
 	\begin{eqnarray}
 		\zeta^{\mu\nu}_{m_J}
 		=\left\lbrace \xi^\mu_{-1}\xi^\nu_{-1},\frac{1}{\sqrt2} \left[\xi^\mu_{-1} \xi^\nu_0+\xi^\mu_0 \xi^\nu_{-1}\right],\frac{1}{\sqrt6} \left[\xi^\mu_{+1} \xi^\nu_{-1}+\xi^\mu_- \xi^\nu_{+1}+2\xi^\mu_0 \xi^\nu_0\right],\frac{1}{\sqrt2} \left[\xi^\mu_{+1} \xi^\nu_0+\xi^\mu_0 \xi^\nu_{+1}\right],\xi^\mu_{+1} \xi^\nu_{+1}\right\rbrace_{m_J} \, ,
 		\nonumber\\\end{eqnarray}
 	with $m_J=-2,\ldots, 2$. For the case $J=3$, we have
 	\begin{eqnarray}
 		\zeta^{\mu\nu\rho}_{m_J}
 		&=&\left\lbrace \xi^\mu_{-1} \xi^\nu_{-1} \xi^\rho_{-1},\frac{1}{\sqrt{3}} \left(\xi^\mu_{-1} \xi^\nu_{-1} \xi^\rho_0 + \xi^\mu_{-1} \xi^\nu_0 \xi^\rho_{-1} + \xi^\mu_0 \xi^\nu_{-1} \xi^\rho_{-1}\right),\right.\nonumber\\&&\left.\frac{1}{\sqrt{15}} \left[2\left(\xi^\mu_{-1} \xi^\nu_0 \xi^\rho_0 + \xi^\mu_0 \xi^\nu_{-1} \xi^\rho_0 + \xi^\mu_0 \xi^\nu_0 \xi^\rho_{-1} \right)+ \xi^\mu_{-1} \xi^\nu_{-1} \xi^\rho_{+1} + \xi^\mu_{-1} \xi^\nu_{+1} \xi^\rho_{-1} +  \xi^\mu_{+1} \xi^\nu_{-1} \xi^\rho_{-1}\right] ,\right.\nonumber\\&&\left.\frac{1}{\sqrt{10}} \left[2\xi^\mu_0 \xi^\nu_0 \xi^\rho_0 + \xi^\mu_{+1}  \xi^\nu_0 \xi^\rho_{-1} + \xi^\mu_0 \xi^\nu_+ \xi^\rho_{-1} +  \xi^\mu_{+1} \xi^\nu_{-1} \xi^\rho_0 + \xi^\mu_{-1} \xi^\nu_{+1} \xi^\rho_0 +  \xi^\mu_{-1} \xi^\nu_0 \xi^\rho_{+1} + \xi^\mu_0 \xi^\nu_{-1} \xi^\rho_{+1}\right],\right.\nonumber\\&&\left.\frac{1}{\sqrt{15}} \left[2\left(\xi^\mu_{+1} \xi^\nu_0 \xi^\rho_0 + \xi^\mu_0 \xi^\nu_{+1} \xi^\rho_0 + \xi^\mu_0 \xi^\nu_0 \xi^\rho_{+1} \right)+ \xi^\mu_{+1} \xi^\nu_{+1} \xi^\rho_{-1} + \xi^\mu_{+1} \xi^\nu_{-1} \xi^\rho_{+1} +  \xi^\mu_{-1} \xi^\nu_{+1} \xi^\rho_{+1}\right],\right.\nonumber\\&&\left.\frac{1}{\sqrt{3}} \left(\xi^\mu_{+1} \xi^\nu_{+1} \xi^\rho_0 + \xi^\mu_{+1} \xi^\nu_0 \xi^\rho_{+1} + \xi^\mu_0 \xi^\nu_{+1} \xi^\rho_{+1}\right),\xi^\mu_{+1} \xi^\nu_{+1} \xi^\rho_{+1}\right\rbrace_{m_J} \, ,
 		\nonumber\\\end{eqnarray}
 	with $m_J=-3,\ldots, 3$.
 \end{widetext}
 
 \subsection{Lorentz-covariant tensors and axial-tensors}~\label{app:Gamma}
 The covariant vertex functions $\Gamma^{\mu\nu\sigma\ldots\omega}$ (for natural parity $P=(-1)^J$) and $\Gamma^{5,\mu\nu\sigma\ldots\omega}$ (for unnatural $P=(-1)^{J+1}$) are symmetric Lorentz and axial tensors of rank $J$, respectively, with $J$ open Lorentz indices $\mu\nu\sigma\ldots\omega$. 
 For natural-parity states, the vertex function $\Gamma^{\mu\nu\sigma\ldots\omega}$ can be expanded in terms of all linearly-independent symmetric Lorentz tensors of rank $J$ that can formed of the available covariants $\gamma^\mu, p^\mu=\hat p_1^\mu+p_2^\mu,$ and $P^\mu=\hat p_1^\mu-p_2^\mu$. For unnatural-parity states, the independent symmetric Lorentz tensors are further multiplied with $\gamma^5$.
 To obtain the {invariant} vertex functions $\Gamma_{m_J}^{(J)}$ and $\Gamma_{m_J}^{5(J)}$, we contract $\Gamma^{\mu\nu\sigma\ldots\omega}$ and $\Gamma^{5,\mu\nu\sigma\ldots\omega}$ with the polarization tensor $\zeta^{\mu\nu\sigma\ldots\omega}_{m_J}$. As a result, any structures that contain one or more $P^\mu$, or structures that include two or more $\gamma^\mu$'s proportional to the metric tensor $\mathrm g^{\mu\nu}$ will be transverse with respect to $\zeta^{\mu\nu\sigma\ldots\omega}_{m_J}$ and therefore do not contribute to the invariant vertex functions.
 
 Thus, the only structures that survive the contraction with $\zeta^{\mu\nu\sigma\ldots\omega}_{m_J}$, for $J>0$, are
 \begin{eqnarray}
 	M^{\mu\nu\sigma\ldots\omega}&=&\frac{1}{J}\left(\gamma^\mu p^\nu p^\sigma\cdots p^\omega+p^\mu\gamma^\nu  p^\sigma\cdots p^\omega\right.\nonumber\\&&+\left.p^\mu p^\nu\gamma^\sigma  \cdots p^\omega+\ldots+p^\mu p^\nu p^\sigma  \cdots \gamma^\omega\right)\,,\nonumber\\
 	\\
 	N^{\mu\nu\sigma\ldots\omega}&=&p^\mu p^\nu p^\sigma \cdots p^\omega\,. 
 \end{eqnarray}    
 For the case $J=0$, the polarization tensor $\zeta_0=\mathbf{1}$ and we are left with a single tensor of rank 0,  which is simply the scalar $N=\mathbf{1}$.
 
 Next, we multiply each of the covariant structures $M^{\mu\nu\sigma\cdots}$ and $N^{\mu\nu\sigma\cdots}$ by the Lorentz scalars $\{\mathbf 1, \Lambda(-p_2)\}$ from the right-hand side and  $\{\mathbf 1, \Lambda(-\hat p_1)\}$ from the left hand side. This operation results in eight covariant tensor terms for $J>0$ (and four terms for $J=0$), each multiplied by a Lorentz-invariant function $G_i=G_i (\hat p_1\cdot p_2,p_2^2)$ with $i=1,\ldots,8$ (and $i=1,\ldots,4$ for $J=0$). 
 In the 1CGE~(\ref{eq:1CSE}), one quark four-momentum is on-shell, i.e., $\hat p_1=m_1^2$. As a result, $\Lambda(\hat p_1)$ appears in~(\ref{eq:1CSE}) and eliminates all terms in the vertex function that are proportional to $\Lambda(-\hat p_1)$, since $\Lambda(\hat p_1)\Lambda(-\hat p_1)=0$. Therefore, only four linearly independent Lorentz structures for $J>0$ (and two for $J=0$) in the half-off-shell 1CGE vertex function. If both quark momenta are off-shell, as in the Bethe-Salpeter approach~\cite{PhysRev.82.291,PhysRev.84.1232} or the four-channel CST~\cite{Savkli:1999me,mass_function_paper,PhysRevD.96.074007}, one has eight Lorentz structures for $J>0$ (and four for $J=0$).     
 Contracting $\Gamma^{\mu\nu\sigma\ldots\omega}$ and $\Gamma^{5,\mu\nu\sigma\ldots\omega}$ with $\zeta^{\mu\nu\sigma\ldots\omega}_{m_J}$  gives the vertex functions of (\ref{eq:tensorGammamunu}) and~(\ref{eq:tensorGamma5munu}), respectively.

 The contractions $\zeta^{\mu\nu\sigma\ldots}_{m_J} M_{\mu\nu\sigma\ldots}$ and $\zeta^{\mu\nu\sigma\ldots}_{m_J} N_{\mu\nu\sigma\ldots}$, reduce in the meson rest frame, where $\boldsymbol P=0$, to the spatial components, i.e.,
 \begin{eqnarray}
 	\zeta^{\mu\nu\sigma\ldots\omega}_{m_J}  M_{\mu\nu\sigma\ldots\omega}&=& (-1)^J p^ip^jp^k\cdots  \gamma^n \zeta^{ijk\ldots n}_{m_J} ,\nonumber\\
 	\zeta^{\mu\nu\sigma\ldots\omega}_{m_J} N_{\mu\nu\sigma\ldots\omega}&=& (-1)^J p^ip^jp^k\cdots p^n \zeta^{ijk\ldots n}_{m_J},\nonumber\\
 \end{eqnarray}
 where the factor $(-1)^J$ arises from the relations $p_i=-p^i$, $\gamma_i=-\gamma^i$, and we have used that $\zeta^{ijk\cdots n}_\lambda$ is totally symmetric in the indices $i, j, k,\ldots n$.
 
 With these manipulations, the vertex functions for natural and unnatural parity states can be written in the meson rest frame as 
 \begin{widetext}
 	\begin{eqnarray}
 		\Gamma_{m_J}^{(J)}(\hat p_1,p_2) &=& (-1)^J \zeta^{ij\ldots m n}_{m_J} p^ip^j\ldots p^m 
 		\left\{G_1\gamma^n   +G_2p^n +
 		\left[G_3\gamma^n  +G_4p^n \right]\Lambda (-p_2)\right\}\,\label{eq:tensorGammamunu2}
 	\end{eqnarray}  
 	and
 	\begin{eqnarray}
 		\Gamma_{m_J}^{5(J)}(\hat p_1,p_2) &=& (-1)^J \zeta^{ij\ldots m n}_{m_J} p^ip^j\ldots p^m 
 		\left\{G_1\gamma^n \gamma^5  +G_2p^n\gamma^5 +
 		\left[G_3\gamma^n\gamma^5  +G_4p^n \gamma^5 \right]\Lambda (-p_2)\right\}\,,\label{eq:tensorGammamunu52}
 \end{eqnarray} \end{widetext}
 respectively. Note that these structures are not orthogonal, but can be orthogonalized by the Gram-Schmidt procedure. The resulting orthogonalized expressions can then be inserted into the 1CGE~(\ref{eq:1CSE}) and the resulting system of equations can, in principle, be solved for the unknown functions $G_1,\ldots,G_4$. This procedure is equivalent to the approach used in this work, where we employ a spherical tensor basis. 
\subsection{Relations between the covariant and spherical-tensor bases}
Here we derive the relations between the invariant functions $G_i$ defined in the covariant tensor basis and the radial partial wave functions $\psi_{j}^\rho (p)$. The derivation here generalizes previous calculations for $J=0$ and $J=1$ to arbitrary values of $J$. 

We start by taking Dirac $\rho$-spinor matrix elements of expressions~(\ref{eq:tensorGammamunu2}) and (\ref{eq:tensorGammamunu52}) to obtain the CST wave functions $\Psi^{+-}$ and $\Psi^{++}$. It is convenient to treat natural- and unnatural-parity mesons separately.
\subsubsection{Natural-parity mesons}
For natural-parity mesons we have
\begin{widetext}
	\begin{eqnarray}
		&&\bar u_1^+(\boldsymbol p,\lambda_1) \Gamma^{(J)}_{m_J}(\hat p_1,p_2)u_2^-(\boldsymbol p,\lambda_2) 
		\nonumber\\	&&\quad=
		(-)^{J+1}\zeta^{ijk\cdots mn}_{m_J} p^ip^j p^k\cdots p^m N_{1 p}N_{2 p}
		\left\lbrace \left(\frac{E_{1 p}+E_{2 p}-\mu}{2m_2} (1-\tilde  p_1\tilde  p_2)G_3+ (1+\tilde  p_1\tilde  p_2)G_1\right) \chi_{\lambda_1}^{\dag}\sigma^n \mathrm i \, \sigma^2\chi^{}_{\lambda_2}\right.\nonumber\\&&\qquad+\left.\hat  p^n\hat  p^l
		\left[  p \left(\frac{E_{1 p}+E_{2 p}-\mu}{2m_2}(\tilde  p_1-\tilde  p_2) G_4-(\tilde  p_1+\tilde  p_2) G_2\right)+ 2\tilde  p_1\tilde  p_2  \left(\frac{E_{1 p}+E_{2 p}-\mu}{2m_2}G_3- G_1\right)\right] \chi_{\lambda_1}^{\dag}\sigma^l \mathrm i \, \sigma^2\chi^{}_{\lambda_2}
		\right\rbrace 
		\nonumber\\\label{eq:GammaTpm1}
	\end{eqnarray}
	and 
	\begin{eqnarray}
		&&\bar u_1^+(\boldsymbol p,\lambda_1) \Gamma^{(J)}_{m_J}(\hat p_1,p_2)u_2^+(\boldsymbol p,\lambda_2)
		\nonumber\\&&\quad=
		(-)^J\zeta^{ijk\cdots mn}_{m_J} p^ip^j p^k\cdots p^m N_{1 p}N_{2 p}
		\left\lbrace \hat  p^n
		\left[  p\left(\frac{E_{2 p}-E_{1 p}+\mu}{2m_2}(1+ \tilde  p_1 \tilde  p_2) G_4+ (1- \tilde  p_1 \tilde  p_2) G_2\right) \right.\right.\nonumber\\&&\qquad+\left.\left. 2\tilde  p_1 \left(\frac{E_{2 p}-E_{1 p}+\mu}{2m_2}G_3+ G_1\right)\right]\chi_{\lambda_1}^{\dag}\mathrm i\, \sigma^2\chi^{}_{\lambda_2} \right.\nonumber\\&&\qquad-\left. \hat  p^l \left(\frac{E_{2 p}-E_{1 p}+\mu}{2m_2} (\tilde  p_1+\tilde  p_2)G_3
		+(\tilde  p_1-\tilde  p_2)G_1\right) \chi_{\lambda_1}^{\dag}\sigma^n \sigma^l\mathrm i \,\sigma^2\chi^{}_{\lambda_2}
		\right\rbrace\,.  \nonumber\\\label{eq:GammaTpp1}
	\end{eqnarray}
\end{widetext}
The two tensor structures 
\begin{eqnarray}
	\zeta^{ijk\cdots mn}_{m_J} p^ip^j p^k\cdots p^m\sigma^n\sigma^2 \label{eq:structureJm1}
\end{eqnarray} and
\begin{eqnarray} 
	\zeta^{ijk\cdots mn}_{m_J} p^ip^j p^k\cdots p^m  \hat  p^n \sigma^2 \label{eq:structureJS0} 
\end{eqnarray} 
appearing in the above expressions are already proportional to $T^{(J-1,1)}_{m_J}$ and $T^{(J,0)}_{m_J}$ of Eqs. (\ref{eq:T1Jm1}) and (\ref{eq:T0J}).  The remaining two structures,    
\begin{eqnarray}
	\zeta^{ijk\cdots mn}_{m_J} p^ip^j p^k\cdots p^m\hat  p^n\hat  p^l\sigma^l \sigma^2
	\label{eq:structureJp1}
\end{eqnarray} 
and 
\begin{eqnarray}
	\zeta^{ijk\cdots mn}_{m_J} p^ip^j p^k\cdots p^m\hat  p^l \sigma^n \sigma^l \sigma^2 \label{eq:structureJS1}
\end{eqnarray} 
are proportional only to the first terms in $T^{(J+1,1)}_{m_J}$ and $T^{(J,1)}_{m_J}$ of Eqs. (\ref{eq:T1Jp1}) and (\ref{eq:T1J}), and are therefore not orthogonal to the tensors (\ref{eq:structureJm1}) and (\ref{eq:structureJS0}) (or, equivalently, to $T^{(J-1,1)}_{m_J}$ and $T^{(J,0)}_{m_J}$). Applying a Gram-Schmidt procedure yields the remaining structures  $T^{(J+1,1)}_{m_J}$ and $T^{(J,1)}_{m_J}$.

An advantage of constructing the spherical tensors directly from irreducible tensor products as in (\ref{eq:TSL}) is that one arrives immediately at the orthogonal basis $\left\lbrace T^{(L,S)}_{m_J}\right\rbrace $ without the need to for a Gram-Schmidt procedure. Apart from this technical difference, both procedures are equivalent.
After reorganizing the terms in Eqs.~(\ref{eq:GammaTpm1}) and~(\ref{eq:GammaTpp1}) such that each contribution is proportional to one of the $T^{(L,S)}_{m_J}$, the one-channel CST wave functions follow directly from Eq.~(\ref{eq:CSTwfs}):
\begin{widetext}
	\begin{eqnarray}
		\Psi^{+-}_{\lambda_1 \lambda_2} (\boldsymbol p)
		&=&
		(-)^{J}\zeta^{ijk\cdots mn}_{m_J} p^ip^j p^k\cdots p^m \frac{N_{12}}{2}
		\left\lbrace \left[\frac{(1-\tilde  p_1\tilde  p_2)G_3}{2m_2} + \frac{(1+\tilde  p_1\tilde  p_2)G_1}{E_{1 p}+E_{2 p}-\mu}\right.\right.\nonumber\\&&+\left.\left.\frac{J}{2J+1}\left(
		p \left(\frac{(\tilde  p_1-\tilde  p_2) G_4}{2m_2}-\frac{(\tilde  p_1+\tilde  p_2) G_2}{E_{1 p}+E_{2 p}-\mu}\right)+2\tilde  p_1\tilde  p_2  \left(\frac{G_3}{2m_2}- \frac{G_1}{E_{1 p}+E_{2 p}-\mu}\right)\right)\right] \chi_{\lambda_1}^{\dag}\sigma^n \mathrm i\, \sigma^2\chi^{}_{\lambda_2}\right.\nonumber\\&&+\left.\frac{1}{2J+1}
		\left[  p \left(\frac{(\tilde  p_1-\tilde  p_2) G_4}{2m_2}-\frac{(\tilde  p_1+\tilde  p_2) G_2}{E_{1 p}+E_{2 p}-\mu}\right)+ 2\tilde  p_1\tilde  p_2  \left(\frac{G_3}{2m_2}- \frac{G_1}{E_{1 p}+E_{2 p}-\mu}\right)\right]\right.\nonumber\\&&\qquad\times\left.\chi_{\lambda_1}^{\dag}\left[(2J+1)\sigma^l\hat  p^l\hat  p^n-J\sigma^n \right] \mathrm i\, \sigma^2\chi^{}_{\lambda_2}
		\right\rbrace 
		\nonumber\\
		&=&\sqrt{(2 J-1)!!}\zeta^{ijk\cdots mn}_{m_J} p^ip^j p^k \cdots p^m \nonumber\\&&\times\chi_{\lambda_1}^{\dag}\left\{\frac{1}{\sqrt{(J-1)!}}\psi_{J-1,1} (p)\sigma^n +\frac{1}{\sqrt{(J+1)!}}\psi_{J+1,1}(p)\left[-(2J+1)\sigma^l\hat  p^l\hat  p^n+J\sigma^n \right]  \right\} \mathrm i\, \sigma^2\chi^{}_{\lambda_2}
		\nonumber\\
		&=&{\sqrt{2}}{\sqrt{4 \pi}}\chi_{\lambda_1}^{\dag}\left[\psi_{J-1,1} (p) T^{(J-1,1)}_{m_J} (\boldsymbol p) +\psi_{J+1,1}(p) T^{(J+1,1)}_{m_J} (\boldsymbol p)\right]\chi^{}_{\lambda_2}
		\label{eq:Psipm}
	\end{eqnarray}
	and 
	\begin{eqnarray}
		\Psi^{++}_{\lambda_1 \lambda_2} (\boldsymbol p)&=&
		(-)^{J}\zeta^{ijk\cdots mn} p^ip^j p^k\cdots p^m \frac{N_{12}}{2}
		\left\lbrace 
		\left[  p\left(\frac{(1+ \tilde  p_1 \tilde  p_2) G_4}{2m_2}+ \frac{(1- \tilde  p_1 \tilde  p_2) G_2}{E_{2 p}-E_{1 p}+\mu}\right) +  \frac{ (\tilde  p_1-\tilde  p_2)G_3}{2m_2}
		+\frac{(\tilde  p_1+\tilde  p_2)G_1}{E_{2 p}-E_{1 p}+\mu}\right] \right.\nonumber\\&&\times\left.\chi_{\lambda_1}^{\dag}\hat  p^n\mathrm i\, \sigma^2\chi^{}_{\lambda_2}-  \left(\frac{ (\tilde  p_1+\tilde  p_2)G_3}{2m_2}
		+\frac{(\tilde  p_1-\tilde  p_2)G_1}{E_{2 p}-E_{1 p}+\mu}\right) \chi_{\lambda_1}^{\dag}\left(\sigma^n \sigma^l\hat  p^l-\hat p^n\right)\mathrm i\, \sigma^2\chi^{}_{\lambda_2}
		\right\rbrace\,.  
		\nonumber\\
		&=&\sqrt{(2 J+1)!!}
		\zeta^{ijk\cdots mn} p^ip^j p^k\cdots p^m 
		\chi_{\lambda_1}^{\dag}\left[\frac{1}{\sqrt{J!}}\psi_{J,0}(p)\hat  p^n-\sqrt{\frac{J}{(J+1)!}}\psi_{J,1}(p)\left( \sigma^l\hat  p^l \sigma^n-\hat p^n\right)\right]\mathrm i\, \sigma^2\chi^{}_{\lambda_2}
		\nonumber\\
		&=& {\sqrt{2}}{\sqrt{4 \pi}}\chi_{\lambda_1}^{\dag}\left[\psi_{J,0} (p) T^{(J,0)}_{m_J}(\boldsymbol p) +\psi_{J,1}(p) T^{(J,1)}_{m_J}(\boldsymbol p)\right]\chi^{}_{\lambda_2}\, ,
		\label{eq:Psipp}
	\end{eqnarray}
	and the relations between the partial waves $\psi$ and the invariant functions $G$ can be read off immediately as
	\begin{eqnarray}
		&&\psi_{J-1,1} (p)=
		(-)^{J} \sqrt{\frac{(J-1)!}{(2 J-1)!!}}
		\frac{N_{12}}{2}
		\left[\frac{\left(1-\frac{\tilde  p_1\tilde  p_2}{2J+1}\right)G_3}{2m_2} + \frac{\left(1+\frac{\tilde  p_1\tilde  p_2}{2J+1}\right)G_1}{E_{1 p}+E_{2 p}-\mu}+\frac{J}{2J+1}
		p \left(\frac{(\tilde  p_1-\tilde  p_2) G_4}{2m_2}-\frac{(\tilde  p_1+\tilde  p_2) G_2}{E_{1 p}+E_{2 p}-\mu}\right)\right]\,, \nonumber\\
		&&	\psi_{J+1,1} (p)=
		(-)^{J+1} \sqrt{\frac{(J+1)!}{(2 J-1)!!}}
		\frac{N_{12}}{2}\frac{1}{2J+1}
		\left[  p \left(\frac{(\tilde  p_1-\tilde  p_2) G_4}{2m_2}-\frac{(\tilde  p_1+\tilde  p_2) G_2}{E_{1 p}+E_{2 p}-\mu}\right)+ 2\tilde  p_1\tilde  p_2  \left(\frac{G_3}{2m_2}- \frac{G_1}{E_{1 p}+E_{2 p}-\mu}\right)\right]\,,\nonumber\\
		&&\psi_{J,0} (p)=(-)^{J} \sqrt{\frac{J!}{(2 J+1)!!}}
		\frac{N_{12}}{2}
		\left[  p\left(\frac{(1+ \tilde  p_1 \tilde  p_2) G_4}{2m_2}+ \frac{(1- \tilde  p_1 \tilde  p_2) G_2}{E_{2 p}-E_{1 p}+\mu}\right) +  \frac{ (\tilde  p_1-\tilde  p_2)G_3}{2m_2}
		+\frac{(\tilde  p_1+\tilde  p_2)G_1}{E_{2 p}-E_{1 p}+\mu}\right]\,,
		\nonumber\\
		&&\psi_{J,1} (p)=(-)^{J+1}
		\sqrt{\frac{J(J+1)!}{(2 J+1)!!}}
		\frac{N_{12}}{2}\left(\frac{ (\tilde  p_1+\tilde  p_2)G_3}{2m_2}
		+\frac{(\tilde  p_1-\tilde  p_2)G_1}{E_{2 p}-E_{1 p}+\mu}\right) \,.
	\end{eqnarray}
	For equal quark masses, $m_1=m_2=m$, this reduces to 
	\begin{eqnarray}
		&&\psi_{J-1,1} (p)=
		(-)^{J} \sqrt{\frac{(J-1)!}{(2 J-1)!!}}
		\frac{1}{2E_p}
		\left[\frac{\left(E_p+m-\frac{E_p-m}{2J+1}\right)G_3}{2m} + \frac{\left(E_p+m+\frac{E_p-m}{2J+1}\right)G_1}{2 E_{ p}-\mu}-\frac{J}{2J+1}
		\frac{2p^2 G_2}{2 E_p-\mu}\right]\,, \nonumber\\
		&&\psi_{J+1,1} (p)=
		(-)^{J+1} \sqrt{\frac{(J+1)!}{(2 J-1)!!}}
		\frac{1}{E_p}\frac{1}{2J+1}
		\left[   -\frac{p^2 G_2}{2 E_p-\mu}+ (E_p-m)  \left(\frac{G_3}{2m}- \frac{G_1}{2E_p-\mu}\right)\right]\,,\nonumber\\
		&&\psi_{J,0} (p)=(-)^{J} \sqrt{\frac{J!}{(2 J+1)!!}}p
		\left[  \frac{ G_4}{2m}+ \frac{ m G_2}{E_p\mu} + 
		\frac{ G_1}{E_p\mu}\right]\,,
		\nonumber\\
		&&\psi_{J,1} (p)=(-)^{J+1}
		\sqrt{\frac{J(J+1)!}{(2 J+1)!!}}
		\frac{p G_3}{2m E_p}\,.
	\end{eqnarray}
	\subsubsection{Unnatural-parity mesons}
	The corresponding calculation of the $\rho$-spinor matrix elements for the vertex function $\Gamma^{5(J)}_{m_J}(\hat p_1,p_2)$ of unnatural-parity mesons yields
	\begin{eqnarray}
		&&\psi_{J,0} (p)=(-)^{J} \sqrt{\frac{J!}{(2 J+1)!!}}
		\frac{N_{12}}{2}
		\left[  p\left(\frac{(1+ \tilde  p_1 \tilde  p_2) G_2}{E_{1 p}+E_{2 p}-\mu}+ \frac{(1- \tilde  p_1 \tilde  p_2) G_4}{2m_2}\right) +  \frac{ (\tilde  p_1-\tilde  p_2)G_1}{E_{1 p}+E_{1 p}-\mu}
		+\frac{(\tilde  p_1+\tilde  p_2)G_3}{2m_2}\right]\,,
		\nonumber\\
		&&\psi_{J,1} (p)=(-)^{J+1}
		\sqrt{\frac{J(J+1)!}{(2 J+1)!!}}
		\frac{N_{12}}{2}\left(\frac{ (\tilde  p_1+\tilde  p_2)G_1}{E_{1 p}+E_{1 p}-\mu}
		+\frac{(\tilde  p_1-\tilde  p_2)G_3}{2m_2}\right)\,,\nonumber\\
		&&\psi_{J-1,1} (p)=
		(-)^{J} \sqrt{\frac{(J-1)!}{(2 J-1)!!}}
		\frac{N_{12}}{2}
		\left[\frac{\left(1-\frac{\tilde  p_1\tilde  p_2}{2J+1}\right)G_1}{E_{2p}-E_{1p}+\mu} + \frac{\left(1+\frac{\tilde  p_1\tilde  p_2}{2J+1}\right)G_3}{2m_2}+\frac{J}{2J+1}
		p \left(\frac{(\tilde  p_1-\tilde  p_2) G_2}{E_{2p}-E_{1p}+\mu}-\frac{(\tilde  p_1+\tilde  p_2) G_4}{2m_2}\right)\right]\,, \nonumber\\
		&&	\psi_{J+1,1} (p)=
		(-)^{J+1} \sqrt{\frac{(J+1)!}{(2 J-1)!!}}
		\frac{N_{12}}{2}\frac{1}{2J+1}
		\left[  p \left(\frac{(\tilde  p_1-\tilde  p_2) G_2}{E_{2p}-E_{1p}+\mu}-\frac{(\tilde  p_1+\tilde  p_2) G_4}{2m_2}\right)+ 2\tilde  p_1\tilde  p_2  \left(\frac{G_1}{E_{2p}-E_{1p}+\mu}- \frac{G_3}{2m_2}\right)\right]	 \,.\nonumber\\	
	\end{eqnarray}
	For equal quark masses, $m_1=m_2=m$, this reduces to 
	\begin{eqnarray}
		&&\psi_{J,0} (p)=(-)^{J} \sqrt{\frac{J!}{(2 J+1)!!}}p
		\left[  \frac{ G_2}{2E_p-\mu}+ \frac{  G_4}{2E_p} + 
		\frac{ G_3}{2E_pm}\right]\,,
		\nonumber\\
		&&\psi_{J,1} (p)=(-)^{J+1}
		\sqrt{\frac{J(J+1)!}{(2 J+1)!!}}
		\frac{p G_1}{(2E_p-\mu) E_p}\,,\nonumber\\
		&&\psi_{J-1,1} (p)=
		(-)^{J} \sqrt{\frac{(J-1)!}{(2 J-1)!!}}
		\frac{1}{2E_p}
		\left[\frac{\left(E_p+m-\frac{E_p-m}{2J+1}\right)G_1}{\mu} + \frac{\left(E_p+m+\frac{E_p-m}{2J+1}\right)G_3}{2m}-\frac{J}{2J+1}
		\frac{2p^2 G_4}{2m}\right]\,, \nonumber\\
		&&\psi_{J+1,1} (p)=
		(-)^{J+1} \sqrt{\frac{(J+1)!}{(2 J-1)!!}}
		\frac{1}{E_p}\frac{1}{2J+1}
		\left[   -\frac{p^2 G_4}{2m}+ (E_p-m)  \left(\frac{G_1}{\mu}- \frac{G_3}{2m}\right)\right]	\,.
	\end{eqnarray}
	
\end{widetext}
  
\bibliography{PapersDB}
\end{document}